\documentclass[11pt]{article}
\usepackage{xcolor}
\usepackage{todonotes}

\PassOptionsToPackage{hypertexnames=false}{hyperref}
\usepackage{jheppub}
\usepackage{amssymb,amsfonts,amsmath,amsthm,lineno}
\usepackage{comment}
\usepackage{enumerate}
\usepackage{mathrsfs}
\usepackage{bbold}
\usepackage{tikz}
\usetikzlibrary{cd,shapes.geometric, arrows}
\usepackage{hyperref}

\makeatletter
\newcommand{\eqnarray@rowlabel}[1]{%
  \Hy@MakeCurrentHrefAuto{equation}%
  \hyper@anchorstart{\@currentHref}\hyper@anchorend
  \eqnarray@originallabel{#1}%
}
\AddToHook{env/eqnarray/begin}{%
  \let\eqnarray@originallabel\label
  \let\label\eqnarray@rowlabel
}
\makeatother
\usepackage{slashed}
\usepackage{color}
\usepackage{empheq}
\usepackage{soul}
\usepackage[normalem]{ulem}

\usepackage{amsmath}
\makeatletter
\newcommand{\Vast}{\bBigg@{4.75}}
\makeatother

\def\a{\alpha}
\def\b{\beta}
\def\g{\gamma}

\def\d{\delta}

\def\e{\epsilon}

\def\m{\mu}
\def\n{\nu}

\def\t{\tau}
\def\th{\theta}

\def\O{\Omega}

\newcommand{\be}{\begin{equation}}
\newcommand{\ee}{\end{equation}}
\newcommand{\bea}{\begin{eqnarray}}
\newcommand{\eea}{\end{eqnarray}}

\newcommand\qt\tau

\renewcommand{\tilde}[1]{\widetilde{#1}}

\newcommand{\del}{\partial}

\makeatletter
\renewcommand{\@seccntformat}[1]{\csname the#1\endcsname.\,\,}
\makeatother
\allowdisplaybreaks

\let \savenumberline \numberline
\def \numberline#1{\savenumberline{#1.}}

\makeatletter
\def\@fpheader{\relax}
\makeatother

\def\bea{\begin{eqnarray}}
\def\eea{\end{eqnarray}}

\def\nn{\nonumber}

\usetikzlibrary{shapes,arrows,chains}
\usetikzlibrary{decorations.markings}
\usetikzlibrary{decorations.pathmorphing}
\usetikzlibrary{shapes.multipart}
\tikzset{snake it/.style={decorate, decoration=snake}}

\usepackage{tikz-cd}
\usetikzlibrary{cd,decorations.markings, arrows.meta}

\newcommand\scalemath[2]{\scalebox{#1}{\mbox{\ensuremath{\displaystyle #2}}}}
\title{Stationary solutions in the small-$c$ expansion of GR}
\author[a]{Enes Bal,}
\author[a]{Ertu\u{g}rul Ekiz,}
\author[a]{Emre Onur Kahya,}
\author[b]{and Utku Zorba}
\emailAdd{bale19@itu.edu.tr}
\emailAdd{ekize15@itu.edu.tr}
\emailAdd{eokahya@itu.edu.tr}
\emailAdd{utku.zorba@iuc.edu.tr}
\affiliation[a]{Department of Physics, Istanbul Technical University \\
Maslak 34469 Istanbul, T\"{u}rkiye \smallskip
}
\affiliation[b]{Department of Engineering Sciences, Istanbul University-Cerrahpaşa\\
Avcilar 34320 Istanbul, T\"{u}rkiye
}
\abstract{We study the small-$c$ expansion of general relativity in ADM variables up to next-to-next-to-leading order (NNLO). We show that, in the stationary sector, this formulation renders the field equations more tractable for explicit solution building. The stationary equations acquire a structure, in which the leading lapse and spatial metric are determined by a closed subsystem, the NLO rotational field satisfies a linear equation on that background, and the NNLO fields satisfy a linear system sourced by the NLO data. The stationary sector exhibits both strong-gravity and weak-field branches, whose structure becomes richer at NNLO. In the strong-gravity branch, we first obtain exact vacuum solutions of NLO Carroll gravity, including the Lense--Thirring and rotating C-metric backgrounds. At NNLO, we then construct the corresponding Lense--Thirring-type and C-metric-type exact vacuum geometries. These solutions also arise from the small-$c$ expansion of the Kerr and rotating C-metric geometries around the strong-gravity background, up to $\mathcal{O}(J)$ at NLO and up to $\mathcal{O}(J^3)$ at NNLO. In the weak-field branch, we find exact Hartle--Thorne-type solutions with an independent quadrupole moment, together with exact spin-squared corrections and a mixed quadrupolar-rotating solution. We further extend the $\ell=0,2$ sector by including higher multipoles up to $\ell=4$, where $\ell$ denotes the multipole index. These results show that the full NLO/NNLO theory admits a richer stationary vacuum sector than the magnetic Carroll truncation. More broadly, the ADM formulation provides a practical framework for constructing and analyzing stationary backgrounds in the small-$c$ expansion of general relativity, and may also offer a useful framework for organizing rotational and higher-multipole deformations motivated by relativistic compact-object geometries.}
\begin{document}

\maketitle
\vfill\eject

\section{Introduction}

The small-$c$ expansion of general relativity leads to Carrollian gravity organized order by order in the parameter $c$ \cite{Hansen:2021fxi}. The resulting theories are referred to as the electric/LO (leading order), NLO (next-to-leading order), and NNLO (next-to-next-to-leading order) sectors of Carroll gravity. Accordingly, a consistent truncation of the NLO theory gives rise to magnetic Carroll gravity. Recently, it was shown that magnetic Carroll gravity does not admit rotating solutions \cite{Kolar:2025ebv}.
However, the corresponding solution spaces of Carroll gravity remain poorly understood, especially beyond the magnetic truncation. This suggests that the full NLO/NNLO theories, which contain a richer structure than the magnetic truncation, may admit nontrivial rotating solutions. In this sense, the small-$c$ expansion may provide an organizational framework analogous to the role played by the large-$c$ expansion in post-Newtonian and multipolar descriptions of slowly rotating and weakly deformed compact objects.

\paragraph{}
Carrollian geometries have been studied in various areas of theoretical physics, ranging from physics on null hypersurfaces and black-hole-horizon dynamics to string theory and holography (see \cite{Bagchi:2025vri,Bergshoeff:2022eog,Blair:2024aqz,deBoer:2021jej,deBoer:2023fnj,Ciambelli:2025unn,Ruzziconi:2026bix,Bagchi:2026wcu,Bagchi:2026qpi,Nguyen:2025zhg,Blair:2025nno} and references therein). Recent work has also applied Carrollian methods to string dynamics near BTZ black holes \cite{Banerjee:2025bkg}. Early studies of Carrollian-like structures originated from the electric limit of GR, first introduced in \cite{Isham:1975ur,Henneaux:1979vn}, in the context of strong gravity, where the gravitational theory exhibits ultralocal behavior. Carrollian geometry was later shown to be an indispensable tool for studying dynamics near spacelike singularities \cite{Oling:2024vmq}, especially in the context of Belinski--Khalatnikov--Lifshitz (BKL) dynamics \cite{Belinsky:1970ew}. A systematic analysis of the small-$c$ expansion of GR was initiated in \cite{Hansen:2021fxi}, and some nontrivial solutions of the electric and magnetic theories were explored with and without matter couplings. Further studies have explored Carrollian black-hole-like solutions and particle dynamics around these geometries using magnetic Carroll gravity and its higher-curvature extensions \cite{Chen:2024how,Kolar:2025ebv,Tadros:2024bev,deBoer:2023fnj,Perez:2021abf,Perez:2022jpr}. Despite this progress, the solution space of the full NLO and NNLO Carrollian theories remains much less understood. In this paper, we investigate this problem in more detail.

\paragraph{}

We show that the small-$c$ expansion of general relativity, which gives rise to LO, NLO, and NNLO Carroll gravity, captures the strong-gravity regime together with rotational corrections up to order $J^3$. In the weak-field regime, it also captures an independently parametrized quadrupolar deformation $Q$
and rotational corrections up to order $J^2$. Moreover, it admits an extension to higher multipoles up to $\ell = 4$. Accordingly, the full NLO and NNLO Carroll theories provide a finite-order framework for organizing exterior rotational and multipolar deformations motivated by compact-object geometries in the strong-gravity and weak-field regimes. As a result, this framework allows the rotational and multipolar structure of stationary geometries to be resolved directly from the field equations, order by order. The resulting hierarchy is recursive, with the solution at each order providing the data entering the equations at the next. One may also view the Carrollian expansion as a perturbative framework for strong-gravity and weak-field backgrounds, incorporating corrections associated with rotation, mass, and higher multipole moments.

\paragraph{}

In order to solve the field equations of the NLO and NNLO actions, we use the ADM decomposition of GR as a starting point \cite{Arnowitt:1962hi}. In this formulation, the full NLO/NNLO Carroll gravity is converted into a constrained stationary system suitable for explicit solution building. In the context of non-Lorentzian gravity, the first 3+1 formulation of nonrelativistic gravity was constructed in \cite{Elbistan:2022plu}, where it was also shown that the KS (Kol--Smolkin) decomposition of GR \cite{Kol:2007bc} is suitable for the large-$c$ expansion (see also the odd expansion in \cite{Ergen:2020yop,Elbistan:2025vyh}). In the Carrollian setting, previous work relates the Carrollian expansion to ADM variables \cite{Tadros:2024uni} and develops Hamiltonian formulations of the electric and magnetic limits \cite{Campoleoni:2022ebj,Henneaux:2021yzg}. In this paper, we first construct the known NLO Carroll gravity of \cite{Hansen:2021fxi} in ADM form.  We then show that, in this formulation, the field equations take a relatively simple form enabling us to solve the NLO and NNLO equations more systematically. One of the crucial features of the ADM decomposition in the Carrollian context is that the Ricci scalar depends only on the spatial hypersurface, and the action resembles the coupling of various fields to three-dimensional Euclidean gravity.

\paragraph{}

Using the ADM form of NLO Carroll gravity in four dimensions, we show that a Lense--Thirring geometry is a vacuum solution of NLO Carroll gravity. To our knowledge, this is the first nontrivial rotating stationary vacuum configuration identified in the full NLO small-$c$ theory. This geometry can also be obtained from the small-$c$ expansion of the Hartle--Thorne, Kerr, and Lense--Thirring geometries by suitable scalings of parameters such as the mass $m$, gravitational constant $G$, and rotation parameter $J$. Since the Hartle--Thorne and Lense--Thirring geometries are approximate slow-rotation exterior metrics rather than exact vacuum solution families of GR such as Kerr, this also shows that approximate relativistic backgrounds can still give rise to exact vacuum solutions of NLO/NNLO Carroll gravity after expansion. More generally, the solutions obtained in this work fall into two distinct regimes, namely the strong-gravity and weak-field branches, at both NLO and NNLO. In the strong-gravity branch, together with the Lense--Thirring solution at NLO, we obtain a slowly rotating C-metric solution. At NNLO, we extend the NLO solutions with $J^3$ contributions for both the Lense--Thirring and C-metric geometries. We then consider the weak-field regime. In this case, apart from the trivial gravitomagnetic solutions at NLO, we obtain nontrivial rotating, quadrupolar, higher-multipole, and accelerating backgrounds around flat space, analogous to those appearing in weak-field nonrelativistic expansions, now in a Carrollian setting. In particular, the weak-field NNLO sector includes an exact Hartle--Thorne-type branch together with its extension to higher multipoles up to $\ell=4$. An important point is that the same Carrollian geometries can also be recovered by directly expanding the corresponding relativistic metrics with appropriate parameter scalings, and the two approaches agree. This opens up the possibility of background solutions in both nonrelativistic and Carrollian settings that do not originate directly from exact relativistic metrics. In particular, there may exist genuinely non-Lorentzian background solutions that cannot be obtained from the scaling or expansion of any relativistic spacetime metric. Likewise, there may exist exact solutions of Carrollian gravity even when the corresponding relativistic metric is not an exact solution of GR. We expect that the ADM form of Carroll gravity and its extensions will be useful for exploring further non-Lorentzian geometries in both the strong-gravity and weak-field regimes. Here an ``exact NLO/NNLO solution'' satisfies all field equations at that finite order, irrespective of whether its relativistic parent is exact, approximate, or absent. Throughout, the parameters $m$, $J$, $Q$ and $\alpha$ are integration constants fixed by matching to the relativistic seed geometries rather than intrinsic Carrollian charges.

More specifically, the strong NLO Lense--Thirring configuration is obtained directly from the NLO equations and agrees with the relevant expansions of exact Kerr and approximate Lense--Thirring/Hartle--Thorne geometries. The strong NNLO and weak Kerr-type branches are inherited from exact Kerr, the C-metric-type branches match the exact rotating C-metric, and the weak Hartle--Thorne-type branch matches the approximate Hartle--Thorne metric. By contrast, the mixed  branch and the higher-multipole solutions through $\ell=4$ are found directly from the NNLO equations; the latter do not arise from the Hartle--Thorne expansion, and no identification with a single exact relativistic family is claimed for either class.

\paragraph{}
It is useful to separate the ingredients of this work according to their status in the literature. The small-$c$ expansion and the covariant electric (LO) and NLO actions were obtained in \cite{Hansen:2021fxi}. Magnetic Carroll gravity arises from the NLO theory through a Lagrange-multiplier truncation; its static backgrounds are known \cite{Hansen:2021fxi,Kolar:2025ebv,deBoer:2023fnj}, while rotating stationary solutions within this truncation are excluded by the no-go result of \cite{Kolar:2025ebv}. The present work provides the ADM formulation of the full NLO theory and its stationary solutions, together with the NNLO action, the NNLO field equations and their stationary reduction, and the structural properties of the stationary sector discussed below.

This work is organized as follows: In Sec.~II, we review the ADM form of general relativity, introduce the small-$c$ expansion of the ADM variables, and summarize the relevant static vacuum backgrounds of magnetic and NLO Carroll gravity. In Sec.~III, we construct stationary solutions in the strong-gravity branch and analyze the exact NLO and NNLO sectors. In Sec.~IV, we turn to the weak-field branch and discuss the corresponding rotating, quadrupolar, and higher-multipole solutions. We conclude in Sec.~V with a summary and outlook. Technical details and complementary derivations are collected in the appendices.


\section{The 3+1 decomposition}
The ADM decomposition is useful because it isolates the constrained structure of the theory and provides a practical framework for solving the field equations. Let us review the general procedure for the 3+1 decomposition outlined in \cite{Elbistan:2022plu}. For a chosen timelike vector $u^\mu$, we introduce the dual one-form $n_\mu$ satisfying
\bea
g_{\mu\nu}\, u^\mu\, u^\nu = -1\,, \qquad n_\mu = g_{\mu\nu}\, u^\nu\,.
\eea 
Then the spacetime metric can be written in the following form:
\bea
ds^2 = - n^2 + \hat \gamma_{ij}\,e^i \, e^j\,,\qquad g^{\mu\nu}\,\partial_\mu\,\partial_\nu = -u^2 + \hat \gamma^{ij}\,e_i\,e_j\,.
\eea 
Here we have $\hat \gamma_{ij} = g_{\mu\nu}\, e^\mu{}_i e^\nu{}_j$ and $\hat \gamma^{ij} = g^{\mu\nu}\, e_\mu{}^i e_\nu{}^j$. The various inverse relations are given by
\bea
e_\mu{}^i \, e^{\mu}{}_j = \delta^i{}_j\,, \qquad e_\mu{}^i\, u^\mu = e^{\mu\, i}\, n_\mu=0\,.
\eea 
In the ADM decomposition, the vectors and vielbeins are defined as follows\footnote{We use hatted variables for the relativistic ADM fields.}:
\begin{align}
&e_i = \partial_i\,,&\qquad  &n = - c\, \hat N\, dt\,, \nn\\
&e^i = dx^i + \hat N^i\,c\,dt\,, &\qquad & u= \hat N^{-1}\, \left(c^{-1}\,\partial_t - \hat N^i\,\partial_i\right)\,.\label{3.4}
\end{align}
Here $u$ is the normal to the spatial hypersurface, $\partial_i$ is the tangent basis vector, and $\hat \gamma_{ij}$ is the induced spatial metric. We keep the speed of light $c$ explicitly for later convenience. Note that for $c \rightarrow 0$, we still have independent vector fields spanning the Carrollian geometric structure. First of all, the metric can be written in the well-known ADM form:
\begin{equation} 
	\label{metcase1}
	ds^2 = -c^2\, \hat N^2dt^2 + \hat \gamma_{ij} (dx^i + c\,\hat N^i dt) (dx^j + c\,\hat N^j dt)\,. 
\end{equation}
Note that the inverse metric is also given by
\begin{equation}
	g^{\m\n}\partial_\m\partial_\n=-c^{-2}\,\hat N^{-2}(\partial_t-c\,\hat N^i\partial_i)^2+ \hat \gamma^{ij}\partial_i\partial_j\,.\label{invmetcase1}
\end{equation}
The ADM action \cite{Arnowitt:1962hi} is then
\begin{equation}
S_{\text{ADM}} = \frac{c^{4}}{16\pi G}
    \int\,d^{d+1}x\, \hat N\,\sqrt{\hat \g}\, \left(\hat R+ \,\hat K_{ij} \hat K^{ij}-\hat K^2\, \right)\,,
  \label{preADM}
\end{equation}
where the extrinsic curvature is defined as
\begin{align}
	\hat K_{ij}&=\hat N^{-1}\left(\frac{1}{2\,c}\partial_t \hat \gamma_{ij}\,- \,\hat\nabla_{(i}\hat N_{j)}\right) \,, \label{Kexplicit}
\end{align}
and $\hat \nabla_i$ is the covariant derivative with respect to the spatial metric $\hat \gamma_{ij}$. In the next subsection, we formulate the small-$c$ expansion in ADM variables order by order. The leading-order action is called \textit{electric Carrollian gravity}, whereas a consistent truncation of the NLO action defines \textit{magnetic Carroll gravity}. Although manifest covariance is lost in this description, the resulting formalism is well suited for constructing solutions order by order in the small-$c$ expansion.
\subsection{Small-\texorpdfstring{$c$}{c} expansion in ADM variables}
We present an even expansion of the ADM variables, following the expansions introduced in \cite{Hansen:2021fxi}\footnote{See Appendix~\ref{covCar} for details.}. We assume that the ADM variables can also be expanded as analytic functions of $c$. However, before introducing the expansion, we comment on the structure of the action and the ambiguity of the expansion procedure. First, let us rewrite the action \eqref{preADM} explicitly:
\begin{equation}
S_{\mathrm{ADM}}
=
\frac{c^4}{16\pi G}
\int d^{d+1}x\,
\hat N\sqrt{\hat \gamma}\,
\left(
\hat R
+
\frac{\hat \gamma^{ikjl}}{\hat N^2}\,\left(
\frac{1}{4c^2}\,\dot{\hat \gamma}_{ij}\,\dot{\hat \gamma}_{kl}
-\frac{1}{c}\,\dot{\hat \gamma}_{ij}\,\hat \nabla_{(k}\hat N_{l)}
+ \hat \nabla_{(i}\hat N_{j)}\,\hat \nabla_{(k}\hat N_{l)}
\right)
\right)\,, \nn
\end{equation}

where $\hat \g^{ikjl} = \hat \g^{ik} \hat \g^{jl}- \hat \g^{ij}\, \hat \g^{kl}$. It can be seen from the structure of the action that it is organized in both odd and even powers of $c$. If we used an odd expansion of the fields and considered the NLO action ($\mathcal{O}(c^4)$), where the curvature first appears at that particular order, then for the stationary background we would obtain exactly the same field equations as in the relativistic form. This suggests that an odd small-$c$ expansion would naturally emphasize backgrounds dominated by time-dependent data, while stationary spatial curvature would only enter at higher order. However, we will not consider this possibility here and leave the odd expansion for future study\footnote{In the context of the large-$c$ expansion of general relativity, the odd expansion is discussed in \cite{Ergen:2020yop,Elbistan:2025vyh}.}.

\paragraph{}

Without rescaling the shift vector, the $\mathcal{O}(c^4)$ action
captures the stationary sector of Einstein gravity, including Kerr.
If instead we rescale $\hat N^i\to c\,\hat N^i$, stationary rotational
data first enter at NNLO. We choose $\hat N^i\to c^{-1}\hat N^i$ to
retain rotational data already at NLO within the even expansion
\eqref{expand}. This choice is consistent with the discussion following
\eqref{3.4}, but it is not uniquely fixed by the Carrollian limit.
Leaving $\hat N^i$ unscaled introduces odd powers of $c$, while
alternative scalings can lead to different finite-order truncations of
the same relativistic ADM action.

After rescaling the shift vector $\hat N^i$, the ADM Lagrangian takes the following form:
\begin{equation}
\mathcal{L}= \frac{c^{4}}{2\,\kappa}\, \hat N\,\sqrt{\hat \g}\, \left[\hat R\,  + \,c^{-2}\,\left(\hat K_{ij} \hat K^{ij}\,- \,\hat K^2\,\right) \right]\,, \label{rescaledevenADM}
\end{equation}
where we define $\kappa = 8\pi G$ and the new extrinsic curvature by
\begin{align}
	\hat K_{ij}&=\hat N^{-1}\left(\frac{1}{2}\partial_t \hat \g_{ij}\,-\,\,\hat \nabla_{(i}\hat N_{j)}\right) \,. \label{Kexplicit_rescaled}
\end{align}
This rescaling of the shift vector $\hat N^i$ allows us to organize the metric and the action in even powers of $c$. After redefining $\hat N^i$, the metric \eqref{metcase1} becomes
 \begin{equation} 
	\label{metcase2}
	ds^2 = -c^2\, \hat N^2dt^2 +\hat \g_{ij} (dx^i + \hat N^i dt) (dx^j + \hat N^j dt)\,,
\end{equation}
and the inverse metric
\begin{equation}
	g^{\m\n}\partial_\m\partial_\n=-c^{-2}\,\hat N^{-2}(\partial_t-\hat N^i\partial_i)^2+\hat \g^{ij}\partial_i\partial_j\,.\label{invmetcase2}
\end{equation}
Let us start with the following even expansion ansatz for the ADM variables:
\bea
\hat\g_{ij} &=& \gamma_{ij} + c^2\, \beta_{ij} + c^4\, \epsilon_{ij}\,, \nn \\
\hat N  &=& N + c^2\,M + c^4\,P\,, \nn \\
\hat N^i &=&N^i + c^2\,A^i\, + c^4\,Z^i\,.\label{expand}
\eea 
Here and from now on, hatted symbols denote the full relativistic ADM fields; $(\gamma_{ij},N,N^i)$, $(\beta_{ij},M,A^i)$, and $(\epsilon_{ij},P,Z^i)$ are their $c^0$, $c^2$, and $c^4$ coefficients.
From this point forward, all spatial indices on the expansion coefficients (such as $\beta_{ij}, L_{ij}$) are raised and lowered using the leading-order spatial metric $\gamma_{ij}$, and $\nabla_i$ denotes the covariant derivative compatible with $\gamma_{ij}$. Then, the expansion of the extrinsic curvature is
\bea
\hat K_{ij} =  K_{ij} + c^2\, L_{ij}\,+ \, c^4\, F_{ij}\,,
\eea
where
\bea 
K_{ij} &=&  N^{-1}\left(\frac{1}{2}\partial_t  \g_{ij}\,-\,\,\nabla_{(i}\ N_{j)}\right)\,, \nn \\
L_{ij}  &=& N^{-1} \left( \frac{1}{2}\, \partial_t\, \beta_{ij} - \nabla_{(i}A_{j)} -\beta_{k(i}\, \nabla_{j)}N^k - \frac{1}{2}\, \, N^k\nabla_k\, \beta_{ij} - \,M\,  K_{ij}\right)\,, \nn \\
F_{ij}&=&N^{-1}\Big[\frac12 \partial_t \epsilon_{ij} -\nabla_{(i}Z_{j)}
-\beta_{k(j}\,\nabla_{i)}A^{k}
-\epsilon_{k(j}\,\nabla_{i)}N^{k} \nn \\
&&\qquad -\frac12 N^{k}\nabla_{k}\epsilon_{ij}
-\frac12 A^{k}\nabla_{k}\beta_{ij}-\big(L_{ij}\,M+K_{ij}\,P\big)
\Big].\label{extadmsecond}
\eea
Therefore, the relativistic Lagrangian can be expanded in the following form:
\bea
\mathcal{L} \to c^{2}\,\mathcal{L} _{LO}  + c^4\, \mathcal{L}_{NLO} + c^6\,\mathcal{L}_{NNLO}\,. \nn 
\eea
The explicit forms of the Lagrangians are\footnote{We give details of the NNLO expansion in Appendix~\ref{NNLOaction}.}:
\bea
\mathcal{L}_{LO} &=& \frac{N\, \sqrt{\gamma}}{2\,\kappa} \left(K_{ij} K^{ij}-K^2\,\right)\,, \label{LO} \\
\mathcal{L}_{NLO} &=& \frac{N\, \sqrt{\gamma}}{2\,\kappa} \left( R  + \left( \frac{M}{N} + \frac{\beta}{2}\right)\, \left(K_{ij} K^{ij}-K^2\, \right)  + 2\,K_{ij}\theta^{ij} \right) \,, \label{NLO}
\eea 
where $\beta = \gamma^{ij}\beta_{ij}$, and we define
\bea
\theta^{ij}= \left(L^{ij} - 	\gamma^{ij}L - \beta^{jk}\,K^{i}{}_k   + \beta^{ij}\,K\right)\,.
\eea 
The relation to the covariant Carroll formulation is summarized in Appendix~\ref{covCar}, where we show that the ADM actions presented here are equivalent to the covariant ones up to boundary terms.
The full field equations are provided in Appendix~\ref{eomfull}. In the main text, we focus on the stationary vacuum sectors and their explicit solutions.

\subsection{Carrollian static backgrounds} \label{staticsec}
In this subsection, we summarize three static vacuum solutions of NLO Carrollian gravity (these are also solutions of magnetic Carroll gravity; see \cite{Kolar:2025ebv,Hansen:2021fxi,deBoer:2023fnj}). We use these static solutions as seeds, effectively \textit{dressing} them to construct stationary solutions order by order through the NNLO small-$c$ expansion of general relativity. The static vacuum solutions are obtained by imposing
\bea
N_i = A_i = M = \beta_{ij} =0 \,, \qquad  \partial_t = 0\,, 
\eea 
in the NLO full field equations (see \eqref{eomexpn}), which yields:
\begin{equation}
\begin{aligned}
R &=& 0\,, \nn \\
N G_{ij}-\nabla_{i}\,\nabla_{j}N +\g_{ij}\, \nabla^2\, N &=& 0\,. \label{staticNLO}
\end{aligned}
\end{equation}
We identify three primary static vacuum solutions:
\begin{subequations} \label{static_seeds}
\begin{align}
\text{\bf{Flat solution:}} \quad & N = 1\,, & \gamma_{ij} &=\text{\rm diag}\left(1\,,r^2\,,r^2\sin^2{\th}\right) \,, \label{flat_sol} \\
\text{\bf{Strong gravity:}} \quad & N = \sqrt{1- \frac{2Gm}{r}}\,, & \gamma_{ij} &= \text{\rm diag} \left( \frac{1}{1- \frac{2Gm}{r}}\,, r^2\,,r^2\sin^2{\th}   \right)\,, \label{strong_grav_sol} \\
\text{\bf{C-metric:}} \quad & N = \frac{\sqrt{Q}}{\Omega}\,, & \gamma_{ij} &= \frac{1}{\Omega^2}\text{\rm diag} \left( \frac{1}{Q}\,, \frac{r^2 }{P} \,, P r^2 \sin^2\theta  \right)\,, \label{cmetric_sol}
\end{align}
\end{subequations}
where $P = 1 + 2\alpha Gm \cos\theta$, $Q = (1 - \alpha^2 r^2)(1 - 2Gm/r)$, and $\Omega = 1 + \alpha r \cos\theta$. Here $\alpha$ and $m$ are the acceleration and mass parameters, respectively.
\paragraph{}
Within the even small-$c$ expansion considered here, one should not expect an exact Kerr solution to appear at finite order. Instead, the NLO and NNLO theories admit distinct stationary sectors whose relation to relativistic rotating metrics is captured order by order. Before constructing these stationary metrics, we simplify the field equations by imposing the stationary ansatz: we demand that the fields be time-independent ($\partial_t = 0$) and set the leading-order shift vector to zero ($N_i=0$) in the NLO and NNLO field equations. Henceforth, we use the field equations given in Appendix~\ref{eomreduce}. In this stationary sector, as we will see, the solutions of the NLO and NNLO field equations naturally organize into two distinct small-$c$ regimes, which we refer to as the strong-gravity and weak-field branches. Thus $N^i=0$ removes only the leading shift:
\eqref{expand} still contains the NLO and NNLO rotational fields
$A^i$ and $Z^i$.

For the stationary reduction, the Hamiltonian constraint and
the spatial equations in \eqref{subNLO} jointly constrain the seed
$(N,\gamma_{ij})$, while in this limit the momentum constraint reads
$\nabla^j[N^{-1}(\nabla_{(i}A_{j)}
-\gamma_{ij}\nabla_k A^k)]=0$ and is linear in $A_i$ on that seed.
It has no counterpart in the magnetic theory, where this vector
structure is truncated away. At NNLO, \eqref{subNNLO} is linear in
$(M,\beta_{ij},Z^i)$ with sources at most quadratic in $A^i$.
The stationary problem is therefore solved recursively, the seed
determining $A_i$ through a linear equation and the two together
sourcing a linear system for the NNLO fields. This structure originates
in the order-by-order expansion rather than in the unexpanded ADM
equations, which are nonlinear in all variables.

\section{Strong-gravity stationary branch}
In this section, we analyze the stationary solutions where the gravitational potential remains significant at the leading order. 
By setting the leading-order shift vector to zero ($N_i=0$), we focus on spacetimes that reduce to static Carrollian backgrounds in the $c \to 0$ limit. This choice allows us to treat rotation as a systematic subleading correction to the strong-gravity vacuum. We use the Hartle--Thorne, Kerr, and C-metric geometries as our relativistic seed backgrounds (see Appendices~\ref{HT}, \ref{Kerr}, and \ref{cmetric} for the full Lorentzian forms and expansion details).

The strong-branch expressions are written in Schwarzschild-like ADM slicings. For the Schwarzschild-based branch, our discussion is restricted to the exterior region $r>2Gm$; for the C-metric branch, it is restricted to the chosen static patch in which $Q>0$ and $\Omega\neq0$. Some NNLO expansion coefficients, including $\beta_{rr}$ in \eqref{NNLOLT}, become singular as $r\to2Gm$, so the order-by-order expansion in this slicing is not uniform near the horizon. We therefore do not claim horizon regularity, an interior extension, or direct access to the BKL regime. Establishing whether these divergences can be removed in a horizon-penetrating Carrollian gauge, or instead signal a breakdown of the chosen small-$c$ expansion, lies beyond the scope of this work.

\subsection{NLO solutions} 

At the NLO level, the stationary reduced equations \eqref{subNLO} admit solutions in which the rotational data are encoded in the subleading shift vector $A^i$. To identify these solutions, we adopt an axisymmetric and stationary ansatz. The existence of the Killing vectors $\partial_t$ and $\partial_\phi$ implies that the metric components depend only on the coordinates $(r, \theta)$. Furthermore, since the frame-dragging effects of a rotating source must align with the direction of rotation, we restrict the shift vector to the azimuthal component, $A_i dx^i = A_\phi(r, \theta) d\phi$. This choice corresponds to a toroidal mode in a spherical-harmonic decomposition.

\subsubsection{Lense--Thirring solution} \label{LTsol}
By applying this ansatz to the NLO field equations \eqref{subNLO}, we obtain an exact vacuum solution that reproduces the Lense--Thirring geometry. The ADM components for this solution are\footnote{From now on, we fix the constant factors in the solutions for the fields by comparing them with the small-$c$ expansion of the corresponding relativistic metrics.}:
\begin{equation} \label{CarrollLT}
\begin{aligned}
N &= \sqrt{1 - \frac{2Gm}{r}}\,, \quad \gamma_{ij}&= \text{\rm diag}\left(\frac{1}{1-\frac{2G m}{r}},r^2,r^2\sin^2{\th} \right)\,, \quad A_{\phi} &= -\frac{2GJ}{r} \sin^2\theta\,. 
\end{aligned}
\end{equation}
A notable feature of the Carrollian framework is that the Lense--Thirring metric, which is only an approximate slow-rotation solution in general relativity, emerges here as an exact vacuum solution of the NLO equations (see Appendix~\ref{LT} for details).  This result shows that the full NLO theory admits a nontrivial rotational vacuum sector, encoded in $A_\phi$, that is absent in the magnetic truncation.

\subsubsection{Rotating C-metric solution}
The framework further allows for the inclusion of acceleration. By dressing the static C-metric vacuum presented in Section~\ref{staticsec} with a rotational shift vector, we find the following NLO solution for an accelerating, rotating source:
\begin{equation}
\begin{aligned}
N &= \frac{\sqrt{Q}}{\Omega}\,, \quad \gamma_{ij}&=\frac{1}{\Omega^2}\text{\rm diag} \left(\frac{1}{Q},\frac{r^2}{P},P r^2\sin^2{\th} \right) \,, \quad  A_{\phi} &= \frac{ J \sin^2\theta}{m\,\Omega^2}  (Q- P)\,,
\end{aligned}
\end{equation}
where 
\begin{equation}
\Omega = 1+\alpha r \cos\theta,
\qquad
P = 1+2\alpha G m \cos\theta,
\qquad
Q = (1-\alpha^2 r^2)\left(1-\frac{2\,G\,m}{r}\right)\,. \label{cmetdef}
\end{equation}
This solution demonstrates that the ADM Carrollian framework is robust enough to handle the complex interplay between frame dragging and uniform acceleration at the NLO level. Note that for vanishing acceleration $\alpha=0$, the solution reduces to the Carrollian Lense--Thirring metric.
\paragraph{}

Furthermore, both solutions can be recovered from a systematic small-$c$ expansion of the corresponding relativistic metrics under the following ``strong'' scaling of the mass $m$, the gravitational constant $G$, and the angular momentum $J$:
\begin{equation}m \to c\, m\,, \qquad G \to c\,G\,, \qquad J \to c^3\,J\,.\label{strongscaling}
\end{equation}
In particular, the Carrollian Lense--Thirring geometry arises from the expansion of either the Kerr or Hartle--Thorne metrics, while the rotating C-metric is recovered from its relativistic rotating counterpart (see Appendices~\ref{HT}, \ref{Kerr}, and \ref{cmetric}). The agreement between these two approaches provides a nontrivial consistency check of our analysis and shows that the stationary ADM formulation offers a robust framework for constructing solutions in Carrollian gravity. Within our ADM formulation, these solutions may be viewed as the static vacuum seeds \eqref{static_seeds} dressed by the nontrivial NLO rotational shift \(A_\phi\).

\subsection{NNLO solutions}

NNLO solutions can be obtained by using \eqref{subNNLO} in the presence of the Killing vectors $\partial_t$ and $\partial_\phi$. Since the NLO solutions contain the nontrivial vector $A_i$, at NNLO we must also turn on $M$, $\beta_{ij}$, and $Z_i$. The basic idea behind the NNLO solutions is that at each order we can use the results of the previous order. For instance, in the strong-gravity regime, we can use either \eqref{strong_grav_sol} or \eqref{cmetric_sol} and dress it with the vector field $A_\phi$. The next step is to use these dressed solutions and turn on the remaining NLO fields (for instance, $M$ and $\beta_{ij}$) and, if necessary, some NNLO fields.

\subsubsection{Lense--Thirring-type solution} \label{LTNNLOSol}

If we choose the base metric to be the Lense--Thirring background \eqref{CarrollLT}, the field equations \eqref{subNNLO} admit the following solution:
\bea
N &=& \sqrt{1 - \frac{2Gm}{r}}\,, \quad \gamma_{ij} =\text{\rm diag}\left( \frac{1}{1 - \frac{2Gm}{r}}, r^2, r^2\sin^2{\th} \right)\,, \quad A_{\phi} = -\frac{2GJ}{r} \sin^2\theta\,, \nn \\
M&
=&
\frac{G\,J^2}{\sqrt{1-\frac{2Gm}{r}}}
\left(
\frac{\cos^{2}\theta}{m r^{3}}
+
\frac{2G}{r^{4}}\sin^{2}\theta
\right)\,, \quad Z_\phi  =\frac{2GJ^3}{m^2r^3}\sin^2\theta\,\left(1+\cos^2{\th}+\frac{2G m}{r}\sin^2{\th} \right)\,, \nn \\
\beta_{ij}
&=&
\frac{J^2}{m^2}\,{\rm diag} \left(\frac{\left(1-\frac{2 Gm}{r}\right)\cos^2{\th}-1}{\left(r-2Gm\right)^2},
\;
\cos^{2}\theta,
\;
\sin^{2}\theta\!
\left(1+\frac{2Gm}{r}\sin^{2}\theta\right)
\right)\,. \label{NNLOLT}
\eea

This exact solution of NNLO Carroll gravity also arises from the small-$c$ expansion of the Kerr metric after the strong scaling of the parameters. The details can be found in Appendix~\ref{Kerr}. It is a finite-order strong-branch geometry matched to the slow-rotation expansion of Kerr through order $J^3$.

We also considered the Hartle--Thorne metric under the condition $Q = J^2/Gm$ together with the strong scaling \eqref{strongscaling}. However, we did not find a consistent NNLO solution. A likely reason is that the Hartle--Thorne metric is only an approximate vacuum solution of the Einstein equations and is truncated at lower order in the slow-rotation expansion. In particular, the NNLO strong-branch solution requires a nontrivial $J^3$ contribution, whereas the Hartle--Thorne metric is not sensitive to this order. In addition, the relativistic Hartle--Thorne expansion involves Legendre functions of the second kind, which exhibit a logarithmic divergence near the horizon at $r = 2Gm$. This may further obstruct a consistent strong-branch NNLO matching. By contrast, the Kerr metric reproduces the required $J^3$ structure and therefore yields a consistent NNLO solution after expansion. In this sense, the Hartle--Thorne truncation is not sufficiently rigid to support the NNLO strong-branch solution within the small-$c$ expansion. A more careful analysis of the Hartle--Thorne metric in the strong-gravity regime with $Q \neq J^2/Gm$ would be very interesting.

\subsubsection{Rotating C-metric-type solution}

If we now choose the Carrollian C-metric as the base geometry, the exact solution of NNLO Carroll gravity is given by
\bea
N &=& \frac{\sqrt{Q}}{\Omega}\,, \quad \gamma_{ij}  = \frac{1}{\Omega^2}\text{\rm diag} \left( \frac{1}{Q}, \frac{r^2}{P} , P r^2 \sin^2\theta \right)\,, \qquad  A_{\phi} = \frac{J \sin^2\theta}{m\,\Omega^2}  (Q - P)\,, \nn \\
\beta_{ij}
&=&
\scalemath{0.85}{\frac{J^2}{m^2\,\Omega^2}\,\text{\rm diag}\left(\frac{Q\cos^2\theta -1 + \alpha^2 r^2}{(Q\,r)^2},
\;
\frac{P-r^2\alpha^2}{P^2}\cos^{2}\theta,
\;
\sin^{2}\theta\!
\left( 2P +(r^2\alpha^2 - P)\cos^2\theta - Q\sin^2\theta  \right)
\right)}\,, \nn \\
M &=&
\frac{GJ^2\cos\theta(\cos\theta-\alpha r)}{Nmr^3\Omega}
-\frac{\alpha^2J^2\sin^2\theta}{2Nm^2\Omega^2}
+\frac{\Omega^2}{2 N P r^2 \sin^2\theta}\,A_\phi^2\,, \nn \\
Z_\phi
&=&
\frac{J^3\sin^2\theta}{m^3r^2\Omega^2}
\left[
\Omega\left(\frac{2Gm}{r}\cos^2\theta-P+1\right)-\alpha^2r^2
\right]
-\frac{\Omega^2}{P\,r^2\sin^2\theta}\,
\beta_{\phi\phi}A_\phi\,, 
\label{NNLOCmet}
\eea
where $Q$, $\Omega$, and $P$ are defined in \eqref{cmetdef}. Once again, for $\alpha=0$ this solution reduces to \eqref{NNLOLT}. One can also show that this solution corresponds to the small-$c$ expansion of the rotating C-metric after the strong scaling of the parameters; see Appendix~\ref{cmetric}. Other possible solutions may arise from the slowly accelerating limit of the C-metric. It would be very interesting to explore this point further.
\section{Weak-field stationary branch}

A second regime is the weak-field regime. This regime starts from the flat solution \eqref{flat_sol} or from the mass-independent leading C-metric background obtained under the weak scaling. The leading geometry is therefore flat, in inertial or accelerated coordinates, and the nontrivial structure appears only through the subleading fields in the small-\(c\) expansion. More precisely, the relativistic parameters scale as \(m\to c^2m\) and \(J\to c^4J\). Thus, the mass contribution drops out of the leading background without imposing \(m=0\); in the accelerated rotating branch, this is taken to be nonzero so that \(J/m\) is well defined. These fields encode the first gravitational corrections around the background and allow us to describe stationary effects such as rotation, acceleration, and quadrupole deformations in a controlled way. In particular, this branch is the natural sector in which the weak gravitational potential is visible most directly, making contact with structures familiar from weak-field and nonrelativistic expansions.
\subsection{NLO solutions}
We present solutions to the reduced NLO field equations \eqref{subNLO}. Although the NLO solutions appear rather simple, they serve as the basis for the NNLO solutions of the field equations \eqref{subNNLO}.
\subsubsection{Weak-field gravitomagnetic solution}
In this regime, we can take the flat metric as the base solution and, using \eqref{subNLO}, obtain the following solution describing pure rotation:
\bea
N &=& 1\,, \qquad \gamma_{ij} =\text{\rm diag}\left(1, r^2, r^2\sin^2{\th}  \right)\,, \qquad A_\phi = -\frac{2GJ}{r}\sin^2\theta\,.
\eea
This solution describes the weak-field rotational sector on a flat background. In particular, the nontrivial gravitational information is encoded in the vector field \(A_\phi\), which plays the role of a Lense--Thirring-type potential in this branch. The physical dynamics associated with this weak-field rotational geometry are closely related to the gravitomagnetic sector of general relativity. Similar backgrounds have been widely studied in the context of the Lense--Thirring effect, where the rotational vector potential gives rise to frame dragging, gyroscope precession, orbital precession, and gravitomagnetic clock effects \cite{Mashhoon:2003ax}.

\subsubsection{Weak-field accelerated gravitomagnetic solution}
For the accelerated solution, we obtain 
\begin{equation}
\begin{aligned}
N &= \frac{\sqrt{Q}}{\Omega}\,, \qquad
\gamma_{ij} = \frac{1}{\Omega^2}\text{\rm diag} \left( \frac{1}{Q},  r^2, r^2 \sin^2\theta \right)\,, \qquad
A_{\phi} = -\frac{\alpha^2 J r^2 \sin^2\theta}{m\,\Omega^2}\,,
\end{aligned}
\end{equation}
where
\begin{equation}
\Omega = 1+\alpha r \cos\theta\,, \qquad
Q = 1-\alpha^2 r^2\,.
\label{cmetdefflat}
\end{equation}
This solution may be viewed as the weak-field accelerated counterpart of the gravitomagnetic branch. This metric describes flat space in accelerated coordinates, while the nontrivial stationary structure is encoded in the shift component \(A_\phi\). In this sense, the solution represents an accelerated weak-field gravitomagnetic geometry rather than a standard asymptotically flat rotational background. Here \(J/m\) is the finite ratio of expansion coefficients fixed by the weak scaling above, not a ratio evaluated after setting \(m=0\).


\subsection{NNLO solutions} 
At NNLO, the weak-field stationary branch acquires a much richer structure. While the NLO sector only captures the rotational, or gravitomagnetic, part of the geometry through the vector field \(A_\phi\), at NNLO one also obtains the scalar field \(M\), the spatial deformation \(\beta_{ij}\), and the vector field \(Z_\phi\). This is the first order at which the weak gravitational potential, spin-squared effects, and quadrupolar and higher-multipole deformations can appear simultaneously. For this reason, the NNLO sector is particularly important for making contact with relativistically motivated weak-field stationary geometries, such as the Hartle--Thorne and Kerr metrics, and for understanding how their multipole structure is encoded in the small-\(c\) expansion.

\noindent $\bullet$~\textbf{Linear sector $A_i=0$:} In this sector, we set $A_i=0$, and the only rotational effect is captured by the toroidal vector $Z_\phi$. Since the background is flat, the field equations in \eqref{subsubNNLO} simplify further:
\begin{align}
\delta N:\quad \, & \nabla_j \nabla_i \beta^{ij} - \nabla_i \nabla^i \beta = 0 \,, \nonumber \\
\delta N^i:\quad \,  & 2\, \nabla^jF_{ij} = 0\,,   \nonumber \\
\delta \gamma^{ij}:\quad \,   & \left(\gamma_{ij}\,\nabla^2 - \nabla_i \nabla_j\right) M + \Big(-\frac12\,\nabla_i\nabla_j\beta
+\nabla_l\nabla_{(i}\beta_{j)}{}^{l}
-\frac12\,\nabla_k\nabla^k\beta_{ij}
\Big)=0\,.   \label{redredeom}
\end{align}

This sector contains several important solutions: the Hartle--Thorne-type solution and its multipolar generalization, the Kerr-type solution, and the mixed-type solution.

\noindent $\bullet$~\textbf{Nonlinear sector $A_i\neq0$:} This sector includes only the C-metric-type solution. The field equations are quadratic, as seen from \eqref{subNNLO}.

\subsubsection{Hartle--Thorne-type solution} \label{HTweak}
 In the weak stationary branch, we impose
  \begin{equation}
  N=1,\qquad \gamma_{ij} =\text{\rm diag}\left( 1, r^2,r^2\sin^2{\th} \right),
  \qquad A_i = 0.
  \end{equation}
  Then the reduced NNLO field equations \eqref{redredeom} become linear in $(M,\beta_{ij},Z_i)$. For axisymmetric stationary configurations, the natural scalar ansatz is
  \begin{equation}
  M(r,\theta)=\sum_{\ell\geq 0} M_\ell(r)\,P_\ell(\cos\theta), \label{Mexp}
  \end{equation}
  where $P_\ell$ are Legendre polynomials. For the toroidal vector sector, we take only an azimuthal component,
  \begin{equation}
  Z_i dx^i = Z_\phi(r,\theta)\, d\phi,
  \end{equation}
  with multipole expansion
  \begin{equation}
  Z_\phi(r,\theta)
  = 
  \sum_{\ell\geq 1} Z_\ell(r)\,\sin\theta\,\partial_\theta P_\ell(\cos\theta). \label{Zexp}
  \end{equation}
  This is the standard axisymmetric toroidal harmonic basis. For the spatial tensor sector, one may begin with a diagonal axisymmetric ansatz
  \begin{equation}
  \beta_{ij}dx^i dx^j
  =
  \beta_{rr}(r,\theta)\,dr^2
  +\beta_{\theta\theta}(r,\theta)\,d\theta^2
  +\beta_{\phi\phi}(r,\theta)\,d\phi^2,
  \end{equation}
  with
  \begin{align}
  \beta_{rr}(r,\theta)
  &=
  \sum_{\ell\geq 0} B^{(rr)}_\ell(r)\,P_\ell(\cos\theta), \nn \\
  \beta_{\theta\theta}(r,\theta)
  &=
  \sum_{\ell\geq 0} B^{(\theta\theta)}_\ell(r)\,P_\ell(\cos\theta), \nn \\
  \beta_{\phi\phi}(r,\theta)
  &=
  \sin^2\theta\sum_{\ell\geq 0} B^{(\phi\phi)}_\ell(r)\,P_\ell(\cos\theta). \label{betexp}
  \end{align}
For the lowest nontrivial multipole sector, one may truncate to $\ell=0,2$. This truncation already contains the monopole sector, the independent quadrupole sector, the toroidal dipole rotational sector, and the spin-squared quadrupolar deformation sector. Hence, one of the important solutions in the weak-field NNLO regime for $\ell=0,2$ is the following:
\bea
N&=&1,\qquad \gamma_{ij} =\text{\rm diag}\left(1, r^2, r^2\sin^2{\th} \right)\,, \nn \\
M&=& -\frac{Gm}{r}+ \frac{GQP_2(\cos{\th})}{r^3}\,,\qquad  Z_\phi=-\frac{2GJ}{r}\sin^2\theta\,, \nn \\
\b_{rr}&=&\frac{2Gm}{r}-\frac{2GQ P_2(\cos{\th})}{r^3}\,, \nn \\
\b_{\th\th}&=&-\frac{2 GQP_2(\cos{\th})}{r}\,,\qquad \b_{\phi\phi}=\b_{\th\th}\,\sin^2{\th} \,.
\eea 
This solution describes the weak-field stationary geometry of a stationary multipolar configuration with an independently parametrized coefficient \(Q\). The field \(M\) contains the usual Newtonian monopole term together with the leading quadrupolar correction, while the vector \(Z_\phi\) encodes the rotational, or gravitomagnetic sector normalized by \(J\) through relativistic matching. The spatial tensor \(\beta_{ij}\) captures the corresponding weak-field deformation of the spatial geometry. In this sense, this branch provides the natural weak-field Hartle--Thorne-type solution within the small-\(c\) expansion. Note that the same metric can be obtained from the weak-field expansion of the Hartle--Thorne geometry with \(Q \neq J^2/(Gm)\), so that the quadrupole moment remains independent of the rotation parameter.

The linearity of the field equations \eqref{redredeom} enables us to investigate further. For instance, the following background is also an exact solution of the NNLO field equations for $\ell=4$:
\begin{equation}
\begin{aligned}
&Z_\phi(r,\th)=\frac{A}{r}\sin^2\theta+\frac{B}{r^3} \sin^2\theta \left(5\cos^2\theta-1\right) \\
&M(r,\th)=-\frac{C}{2 r}-\frac{D}{2r^3}P_2(\cos{\th})-\frac{E}{2r^5}P_4(\cos{\th})\\
&\b_{rr}(r,\th)=\frac{C}{r}+ \frac{D}{r^3}P_2(\cos{\th})+\frac{E}{r^5}P_4(\cos{\th})\\
&\b_{\th\th}(r,\th)=\frac{D}{r}P_2(\cos{\th})+\frac{E}{r^3}P_4(\cos{\th})\\
&\b_{\phi\phi}(r,\th)=\frac{D}{r}\sin^2{\th}P_2(\cos{\th})+\frac{E}{r^3}\sin^2{\th}P_4(\cos{\th})\\ \label{octupole}
\end{aligned}
\end{equation}
where $A,B,C,D$, and $E$ are arbitrary constants. Note that this solution, as well as further higher-multipole solutions, cannot be obtained from the expansion of the Hartle--Thorne metric, since the Hartle--Thorne geometry is sensitive only to the quadrupole moment. We expect the general axisymmetric solution to admit expansions of the forms \eqref{Mexp}, \eqref{Zexp}, and \eqref{betexp}. These solutions provide higher-multipole extensions of the Hartle--Thorne-type sector, allowing for more general axisymmetric deformations of astrophysical interest.

\subsubsection{Kerr-type solution}
The weak-field solution with quadratic rotation terms is an exact solution of the NNLO theory:
\bea 
N&=&1,\qquad \gamma_{ij} =\text{\rm diag}\left(1, r^2, r^2\sin^2{\th} \right)\,,\nn \\ 
M&=& -\frac{Gm}{r}\,,\qquad  Z_\phi=-\frac{2GJ}{r}\sin^2\theta\,,\nn \\
\b_{rr}&=&\frac{2Gm}{r} - \frac{J^2}{m^2\,r^2} \sin^2{\th}\,,\nn \\
\b_{\th\th}&=& \frac{J^2}{m^2} \cos^2{\th}\,,\qquad\b_{\phi\phi}=\frac{J^2}{m^2} \sin^2{\th}\,.  \label{weakKerrNNLO}
\eea 
This solution represents the weak-field Kerr-type branch of the NNLO theory. The field \(M\) contains the Newtonian monopole potential, while \(Z_\phi\) encodes the gravitomagnetic sector associated with the angular momentum \(J\). In contrast to the NLO rotational solution, the NNLO tensor \(\beta_{ij}\) contains terms of order \(J^2\), which describe the leading spin-squared deformation of the spatial geometry. In this sense, the solution captures the first weak-field imprint of the Kerr multipole structure within the small-\(c\) expansion. Unlike the Hartle--Thorne-type branch, there is no independent quadrupole parameter here; rather, the quadrupolar deformation is fixed by the rotation, as expected for a Kerr-type geometry \cite{Hansen:1974zz}.
\subsubsection{Mixed-type solution} \label{mixedsec}
The next nontrivial solution is given by  
\bea 
N&=&1,\qquad \gamma_{ij} =\text{\rm diag}\left( 1, r^2, r^2\sin^2{\th} \right) \,,\nn \\ 
M&=& -\frac{Gm}{r} + \frac{GQP_2(\cos{\th})}{r^3}\,,\qquad  Z_\phi=-\frac{2GJ}{r}\sin^2\theta\,,\nn \\
\b_{rr}&=&\frac{2Gm}{r} -\frac{2GQ P_2(\cos{\th})}{r^3} - \frac{J^2}{m^2\,r^2} \sin^2{\th}\,,\nn \\
\b_{\th\th}&=&  -\frac{2 GQP_2(\cos{\th})}{r}+\frac{J^2}{m^2} \cos^2{\th}\,,\nn \\
\b_{\phi\phi}&=& -\frac{2 GQ\,P_2(\cos{\th})}{r}\sin^2\theta + \frac{J^2}{m^2} \sin^2{\th}\,.  \label{weakmixed}
\eea 
This is an exact solution of NNLO Carroll gravity even if $J=0$ or $Q=0$. For a detailed analysis of this metric and its relativistic uplift, see Appendix~\ref{Mixed}.

This solution represents a weak-field stationary geometry that combines the three main relativistically motivated structures of the NNLO branch: the Newtonian monopole, the gravitomagnetic rotational sector, and an independent quadrupole deformation. The field \(M\) contains both the monopole term and the leading quadrupolar correction, while \(Z_\phi\) encodes the frame-dragging sector associated with the angular momentum \(J\). The tensor \(\beta_{ij}\) contains both the quadrupolar spatial deformation and the leading spin-squared corrections proportional to \(J^2\). In this sense, the solution interpolates between the Hartle--Thorne-type and Kerr-type branches. It is therefore closer in spirit to quasi-Kerr or Kerr-like quadrupolar geometries than to the pure Kerr family, since the quadrupole moment remains independent rather than being fixed by the Kerr relation \cite{Mashhoon:2003ax,Hansen:1974zz,Glampedakis:2005cf,Frutos-Alfaro:2015gaa}. Unlike the pure Kerr-type branch, this solution allows the spin-induced and independently parametrized quadrupolar deformations to be disentangled. This type of solution can be generalized to the higher-multipole expansion illustrated in \eqref{octupole}.

\subsubsection{C-metric-type solution}
This solution belongs to the nonlinear sector of the field equations \eqref{subNNLO}. We then obtain the following weak-field exact solution of the NNLO field equations: 
\bea 
N&=&\frac{\sqrt{1-\alpha^2 r^2}}{\Omega}\,,\quad 
\g_{ij}=\frac{1}{\O^2}\text{\rm diag}\left( \frac{1}{1-\a^2r^2}, r^2, r^2\sin^2{\th} \right)\,, A_\phi=-\frac{\alpha^2 J r^2\sin^2\theta}{m\,\Omega^2}\,,  \nn
\\[6pt]
M&=&-N\left(\frac{Gm}{r}+\frac{\alpha^2 J^2\sin^2\theta}{2m^2}\right)\,,\quad Z_\phi
=r^2\sin^2\theta
\left[
-\frac{2GJ(1-\alpha^2 r^2)}{r^3\,\Omega}
+\frac{\alpha^4 J^3}{m^3\Omega^2}
\right] \nn
\\[6pt]
\beta_{rr}
&=&
\frac{1}{(1-\alpha^2 r^2)\,\Omega^2}
\left(
\frac{2Gm}{r}
-\frac{J^2\sin^2\theta}{m^2 r^2}
\right)\,,\nn 
\\[6pt]
\beta_{\theta\theta}
&=&
\frac{r^2}{\Omega^2}
\left(
-2\alpha Gm \cos\theta
+\frac{J^2}{m^2r^2}\cos^2\theta\,(1-\alpha^2 r^2)
\right)\,, \nn 
\\[6pt]
\beta_{\phi\phi}
&=&
\frac{r^2\sin^2\theta}{\Omega^2}
\left(
2\alpha Gm\cos\theta
+\frac{J^2}{m^2 r^2}(1+\alpha^2 r^2)
\right)\,.\label{weakcmetNNLO}
\eea 
This solution may be viewed as the weak-field accelerated Kerr-type branch of the NNLO theory. The parameter \(\alpha\) inherits its acceleration label from matching to the relativistic C-metric. Accordingly, this geometry combines the weak Newtonian potential, frame dragging, and spin-squared deformation in an accelerated background. The field \(M\) contains both the monopole contribution and an acceleration-dependent rotational correction, while \(Z_\phi\) encodes the corresponding gravitomagnetic sector. The tensor \(\beta_{ij}\) then captures the leading spatial deformation induced jointly by rotation and acceleration. Since the solution reduces to the Kerr-type background in the limit \(\alpha\to0\), it can be regarded as the accelerated deformation of the weak-field Kerr background \eqref{weakKerrNNLO}. In this sense, the solution provides a natural weak-field arena for studying how acceleration modifies Kerr-like multipole structure and frame-dragging effects.

\paragraph{}
These solutions can be obtained from the relativistic metrics by the following parameter scalings
       \[
m \;\to\; c^{2}\,m\,,
\qquad 
G \;\to\; c^2\,G\,,
\qquad 
J \;\to\; c^4\,J\,,
\qquad 
Q \;\to\; c^2\,Q\,. 
\]
These scalings correspond to the weak-field regime, as they describe an expansion around the flat solution. This regime is reminiscent of the nonrelativistic expansion, and the Newtonian potential is more transparent in this case (see Appendices~\ref{HT}, \ref{Kerr}, and \ref{cmetric}). It would be interesting to explore this regime in nonrelativistic gravity through the large-$c$ expansion and to clarify its relation to the small-$c$ expansion. We leave this study for future work.

\section{Conclusions} 

In this paper, we formulated the consistent small-$c$ expansion of general relativity in ADM form up to NNLO, both at the level of the action and at the level of the field equations. After rescaling the shift vector $N^i$, the ADM formulation renders the stationary field equations more tractable and allows us to construct explicit vacuum solutions with relatively few assumptions. These results show that the full NLO/NNLO theory admits a richer stationary vacuum sector than the magnetic Carroll truncation, including nontrivial rotational and multipolar structure.

The leading-order theory corresponds to electric Carroll gravity, while a consistent truncation of the NLO theory gives magnetic Carroll gravity. We also provided the map showing the equivalence of our ADM formulation with the covariant Carroll expansion of general relativity up to NLO \cite{Hansen:2021fxi}. The NLO and magnetic sectors admit three static vacuum solutions: flat, strong-gravity, and C-metric backgrounds. Thus, we used them as seed solutions and dressed them with nontrivial rotational and quadrupolar contributions. In the strong-gravity sector, we obtained the Lense--Thirring and rotating C-metric solutions at NLO, while at NNLO we found the Kerr expansion up to order $J^3$ together with the corresponding rotating C-metric expansion. In the weak-field sector, we obtained Hartle--Thorne-type solutions with an independent quadrupole ($\ell=2$) and higher-multipole $(\ell>2)$ moments, as well as Kerr-type and C-metric-type contributions at order $J^2$. We also found a mixed solution containing both $J^2$ and an independent quadrupole parameter $Q$. Moreover, we showed that these Carrollian backgrounds can also be recovered from the small-$c$ expansion of the corresponding relativistic metrics, provided one chooses the appropriate scalings of the parameters $m$, $G$, $J$, and $Q$. In this way, the stationary sector naturally organizes into strong-gravity and weak-field branches.

The labels $m$, $J$, $Q$, and $\alpha$ used in this work are fixed by matching to the corresponding relativistic seed metrics. In the absence of a specification of Carrollian boundary conditions and a construction of the associated charges, we do not identify $m$ and $J$ here as intrinsic conserved charges, nor $Q$ and $\alpha$ as intrinsically defined quadrupole and acceleration parameters. Likewise, the present analysis does not establish horizon data, thermodynamic quantities, or gauge-invariant observables for these backgrounds. These questions are logically separate from the statement that the displayed fields solve the finite-order NLO or NNLO equations and are left for future work.

There are several natural directions for future work. It would be interesting to extend the analysis to higher multipole moments $(\ell>4)$ and to study whether the ADM formulation also reveals genuinely non-Lorentzian stationary backgrounds that do not arise from the expansion of relativistic metrics. It would also be worthwhile to investigate matter-coupled theories and the asymptotic charges associated with the backgrounds constructed in this work. More broadly, the ADM formulation of the small-$c$ expansion provides a useful framework for exploring further rotating and multipolar solutions in Carroll gravity.

It would also be worthwhile to investigate whether our formalism can be used to study tidal perturbations and Love numbers, including in non-asymptotically flat settings such as the Schwarzschild--de Sitter background recently analyzed in \cite{Altas:2026iqc}. In this context, it would be particularly interesting to develop a matter-coupled Carrollian gravity framework and explore whether it can describe the corresponding interior solutions of compact objects. These applications lie beyond the present vacuum construction:
calculating Love numbers requires a perturbation analysis and boundary
data, while modeling compact-object interiors requires a matter action
and matching conditions.

\section{Acknowledgements}
We would like to thank Mehmet Özkan, Oğuzhan Kaşıkçı, and Niels A. Obers for their useful comments and feedback. This work was supported by Scientific Research Projects Department of Istanbul Technical University. Project Number: 48548.  E.O.K. is supported by the Scientific and Technological Research Council of Türkiye (TÜBİTAK) under Grant No. 126F212. U.Z. is supported by the Scientific and Technological Research Council of Türkiye (TÜBİTAK) under Grant Nos.~125F467 and 125F024.

\appendix
\section{Review: Carroll expansion of general relativity} \label{covCar}

In this appendix, we briefly review the small-$c$ expansion of general relativity, first analyzed systematically in \cite{Hansen:2021fxi}. The main assumption underlying the small-$c$ expansion is that the metric components are analytic functions of $c$, which allows one to expand the metric in powers of $c$. The resulting geometry is non-Lorentzian and has a Newton--Cartan-like structure. The higher-order fields capture successive orders in the gravitational interaction, in a way analogous to the large-$c$ expansion, where subleading terms encode strong-gravity effects \cite{VandenBleeken:2017rij, Hansen:2019pkl, Hansen:2020pqs}. The small-$c$ expansion proceeds as follows. First, one introduces the pre-ultra-local (PUL) form of the metric and inverse metric:
\bea
g^{\mu\nu}\, \partial_{\mu}\, \partial_{\nu}\, &=& (- \frac{1}{c^2}\, T^{\mu}\, T^{\nu} + \Pi^{\mu\nu})\,  \partial_{\mu}\, \partial_{\nu}\,, \nn \\
g_{\mu\nu} &=& - c^2\, T_\mu \, T_\nu + \Pi_{\mu\nu}\,. \label{carrollexp}
\eea 
Then one can immediately use the relation $g_{\mu\nu}\, g^{\nu\rho} = \delta^\rho_\mu$ and the PUL decomposition to obtain the following inverse relations:
\bea
T_\mu\ T^\mu &=& - 1\,, \qquad \Pi_{\mu\nu}\, T^{\mu} =  \Pi^{\mu\nu}\, T_{\mu}=0\,, \nn \\
\delta^\mu_\nu&=& -T_\nu\ T^\mu  + \Pi_{\nu\rho}\, \Pi^{\rho\mu}\,.  \label{incvreltpre}
\eea 
Then, the small-$c$ expansion of the PUL variables is
\bea
T^{\mu} &=& v^\mu + c^2 M^\mu\,, \nn \\
\Pi^{\mu\nu} &=& h^{\mu\nu} + c^2 \, \Phi^{\mu\nu}\,, \nn \\
T_{\mu} &=& \tau_\mu + c^2 m_\mu\,, \nn \\
\Pi_{\mu\nu} &=& h_{\mu\nu} + c^2 \, \Phi_{\mu\nu}\,. \label{expCarr}
\eea 
Note that the leading-order geometric objects $(v^\mu,h_{\mu\nu})$ are precisely those that appear in Carrollian geometry and define the underlying non-Lorentzian structure. At leading order, the inverse relations in \eqref{incvreltpre} reduce to
\bea
\tau_\mu\ v^\mu &=& - 1\,, \qquad h_{\mu\nu}\, v^{\mu} =  h^{\mu\nu}\, \tau_{\mu}=0\,, \nn \\
\delta^\mu_\nu&=& -\tau_\nu\ v^\mu  + h_{\nu\rho} h^{\rho\mu}\,,
\eea
At the next order, we also have the inverse relations
\bea
m_\mu\, v^\mu + \tau_\mu\, M^\mu=0\,, \qquad m_\mu\, h^{\mu\nu} + \tau_\mu\, \Phi^{\mu\nu} = 0\,, \qquad
M^\mu\, h_{\mu\nu} + v^\mu \, \Phi_{\mu\nu} = 0\,,
\eea 
from which the $m_\mu$ and $\Phi_{\mu\nu}$ fields can be expressed as 
\bea
m_\mu &=& \tau_\mu\,\tau_\nu\, M^\nu - h_{\mu\nu}\, \tau_{\sigma}\, \Phi^{\nu\sigma}\,, \nn \\
\Phi_{\mu\nu} &=& - h_{\mu\rho}\, h_{\nu\sigma}\, \Phi^{\rho\sigma} + \tau_\mu\, h_{\nu\rho}\, M^{\rho} + \tau_\nu\, h_{\mu\rho}\, M^{\rho}\,. 
\eea
The metric expansion can also be used to expand the Einstein--Hilbert action. The Einstein--Hilbert action is
\bea
S= \frac{c^3}{16\pi G}\, \int\, d^{d+1}\,x\, \sqrt{-g}\, \mathcal{R}\,, \label{eq:SEH-start}
\eea
where $x^\mu = (t, x^i)$ and $\mathcal{R}$ is the $(d+1)$-dimensional Ricci scalar. Equivalently, this can be written as
\begin{equation}
  S_{\text{EH}} = \frac{c^{4}}{16\pi G}
    \int\,d^{d+1}x\, E\, \mathcal{R}\,,
  \label{eq:SEH-with-E}
\end{equation}
where $E = c^{-1}\,\sqrt{-g}$. Using the Carrollian expansion of the metric \eqref{expCarr}, the LO and NLO Lagrangians are obtained in the following forms, respectively:
\bea
\overset{(0)}{L}_{Carr}&=& e\, \left( K^{\mu\nu}\, K_{\mu\nu} - K^2\right)\,  \nn \\
\overset{(2)}{L}_{Carr} &=& e\, \bigg(h^{\mu\nu}\, R_{\mu\nu} + 2 G_{\mu}\, M^{\mu} +  G_{\mu\nu}\, \Phi^{\mu\nu} \bigg)\,, \label{CarrAction2}
\eea 
where $e= \text{det}\left( -\tau_\mu\,\tau_{\nu} + h_{\mu\nu}\right)$, and the LO field equations associated with the NLO fields are given by
\bea
G_{\mu} &=& -\frac{1}{2}\, \tau_{\mu}\, \left(K^{\rho\sigma}K_{\rho\sigma} - K^2 \right) + h^{\nu\rho}\, \nabla_{\rho}\, \left( K_{\mu\nu} - h_{\mu\nu}\, K\right)\,, \nn \\
G_{\mu\nu} &=& - \frac{1}{2}\, h_{\mu\nu}\, \left(K^{\rho\sigma}K_{\rho\sigma} - K^2 \right)  + K \left(K_{\mu\nu} - h_{\mu\nu}\,K \right)  - v^\rho\, \nabla_{\rho}\, \left(K_{\mu\nu} -h_{\mu\nu} \, K \right) \,.  \label{fieldeq1}
\eea 
This is the basic mechanism underlying the magnetic truncation. The LO equations are always obtained by varying the NLO action with respect to the NLO fields \cite{Hansen:2020pqs}.
The metric compatibility conditions are
\bea
\nabla_\mu\, v^\nu&=& 0\,, \nn \\
\nabla_\mu\, h_{\nu\rho} &=&0 \,,
\eea 
where the metric-compatible connection is
\bea
\Gamma^{\rho}_{\mu\nu} = - v^{\rho}\, \partial_{(\mu}\tau_{\nu)} - v^\rho\, \tau_{(\mu}\,{\mathcal{L}}_{v}\, \t_{\nu)} + \frac{1}{2}\, h^{\rho\lambda}\, \left( \partial_\mu h_{\nu\lambda} + \partial_\nu h_{\mu\lambda} -\partial_\lambda h_{\mu\nu}   \right) - h^{\rho\lambda}\, \tau_{\nu}\, K_{\mu\lambda}\,.
\eea 
Here, the extrinsic curvature is defined by $K_{\mu\nu} = -\frac{1}{2}\, \mathcal{L}_v\, h_{\mu\nu}$, and it satisfies the projection condition
\bea
v^\mu\, K_{\mu\nu} = 0\,.
\eea 
This connection satisfies the following total derivative identity for an arbitrary tensor $X^\mu$:
\bea
\partial_\mu \left(e\, X^\mu\right) = e\left( \nabla_\mu X^\mu  + \tau_\mu\, X^\mu\, K \right)\,. 
\eea 
The LO and NLO actions are invariant under the following local boost transformations:
\bea
&\delta \tau_\mu = \lambda_\mu\,, \qquad &\delta h^{\mu\nu} = 2\, \lambda^{(\mu}\,\tau^{\nu)}\,,\nn\\
&\delta v^\mu = 0\,, \qquad &\delta h_{\mu\nu} =0\,, \label{carrtrans}
\eea 
and the NLO fields transform as
\bea
\delta M^\mu = \lambda^\mu\,, \qquad \delta\,\Phi^{\mu\nu} = 2\,\lambda^{(\mu}\, M^{\nu)}\,. \label{carrtrans2}
\eea 
where $v^\mu\,\lambda_\mu=0$. Note that the metric determinant $e$ is invariant under Carroll boosts since
\[
\delta e = e\left(\tau_\mu\,\delta v^\mu - \frac{1}{2}\,h_{\mu\nu}\,\delta h^{\mu\nu}\right).
\]
Note that $K_{\mu\nu}$ is also a Carroll-invariant object.
\paragraph{Magnetic action:}
The consistent truncation of the NLO fields leads to the magnetic action
\bea
\mathcal{L}_{MC} = e\, \left(h^{\mu\nu}\, R_{\mu\nu}\,  + \phi^{\mu\nu}\,K_{\mu\nu}\, \right)\,, \label{covmag}
\eea
where $\phi^{\mu\nu}$ is a Lagrange multiplier that imposes the Hamiltonian constraint $K_{\mu\nu}=0$; see \cite{Hansen:2021fxi} for details. For the magnetic action to be Carroll-invariant, the Lagrange multiplier must transform as
\bea
\delta\, \phi^{\mu\nu} = 2 \left( 
    - h^{\mu \nu} \lambda^\rho \tau_{\rho \sigma} v^\sigma 
    + h^{\rho (\mu} \lambda^{\nu)} \tau_{\rho \sigma} v^\sigma 
    - h^{\rho (\mu} \nabla_\rho \lambda^{\nu)} 
    + h^{\mu \nu} \g^\rho_{\ \sigma} \nabla_\rho \lambda^\sigma 
\right)\,. 
\eea 
The magnetic action captures most static solutions of GR, whereas the electric action captures the full time-dependent solutions of GR, as can be seen from the structure of the field equations in ADM Carroll form.

\paragraph{Covariant Carroll gravity vs.\ ADM Carroll gravity:}
The covariant LO and NLO Carroll gravity actions can be decomposed into ADM form. This decomposition is useful for solving the field equations up to NNLO. The LO and NLO actions can be mapped to those of \cite{Hansen:2021fxi} by using the following map for the PUL variables:
 \bea
T^\mu &=& \hat N^{-1}\, \left(\delta^\mu_0 -\delta^\mu_i \, \hat N^i\right)\,, \nn \\
\Pi^{\mu\nu} &=& \delta^\mu_i\delta^\nu_j\, \hat \gamma^{ij}\,. 
\eea 
After using the even-power expansion, we obtain the Carrollian map
\bea
v^\mu &=&  N^{-1}\, \left(\delta^\mu_0 -\delta^\mu_i \, N^i\right)\,, \nn \\
h^{ij} &=&\gamma^{ij}\,, \nn \\
M^\mu &=&  N^{-1}\, \left(-\frac{M}{N}\,\delta^\mu_0 +  \delta^\mu_i \, (\frac{M}{N}\,N^i -  A^i ) \right)\,\nn \\
\Phi^{ij} &=& - \beta^{ij}\,.  \label{map1}
\eea 
From this map one also obtains the following component relations:
\bea
h_{ij}&=& \gamma_{ij}\,, \quad h_{0i} = \gamma_{ij}\, N^j\,, \quad h_{00} = \gamma_{ij}\, N^i\, N^j\,, \nn \\
\tau_\mu &=& - N\, \delta^0_\mu\,,  \nn \\
m_{\mu}&=& - M\, \delta^0_\mu\,,  \nn \\
\Phi_{ij} &=& \beta_{ij}\,, \quad \Phi_{i0} = \beta_{ij}\,N^j + \gamma_{ij}\, A^j\,,\quad  \Phi_{00} = \beta_{ij}\, N^iN^j + \gamma_{ij}\, A^i\,N^j + \gamma_{ij}\, N^i\, A^j \,.  \label{map2}
\eea 
Therefore, using the maps \eqref{map1} and \eqref{map2}, one can easily show that the actions \eqref{LO} and \eqref{NLO} correspond to the actions in \cite{Hansen:2021fxi} up to a boundary term:
\bea
 B =\frac{1}{\sqrt{\gamma}}\, \partial_t(\beta^{ij}\, \tilde{K}_{ij}\sqrt{\gamma})  - \gamma^{ij}\, \nabla_i \nabla_jN \, + \,\nabla_k (\tilde{K}_{ij}\, \beta^{ij}\, N^k)- 2\, \nabla^j(\tilde{K}_{ij}\,A^i)\,,\label{totalder}
\eea where we define 
\bea
\tilde{K}_{ij} = K_{ij} - \gamma_{ij}\,K\,.
\eea 

Here we assume fixed endpoint data and falloff conditions for which the boundary term and its first variation vanish. The bulk equivalence alone therefore does not imply an equivalence of boundary charges, and the boundary structure of Carrollian geometry requires further examination; see \cite{Ciambelli:2025unn} for a detailed analysis.

\paragraph{Local boosts:}
Let us briefly discuss local Carrollian boost transformations of the ADM variables. Although local boost invariance is manifest in the covariant LO and NLO actions of \cite{Hansen:2021fxi}, the corresponding transformations of the ADM variables are not directly inherited after imposing the ADM gauge choice. In particular, after the decomposition one no longer has a simple, well-defined realization of Carroll boosts on the ADM fields. Following the discussion around Sec.~2.5 of \cite{Hansen:2021fxi,Baiguera:2022lsw}, the notion of a spatial hypersurface is not boost-invariant in Carrollian geometry. This is related to the fact that $\tau_\mu$ transforms under Carroll boosts, unlike in Newton--Cartan geometry, where it is boost-invariant and spatial hypersurfaces can be defined unambiguously. The non-invariance of $\tau_\mu$ allows one to choose a preferred frame locally, and this may be viewed as the choice of an Ehresmann connection. The ADM decomposition automatically selects such a preferred frame through the gauge choice $\tau_i=0$, often referred to as the time gauge in Carrollian gravity \cite{Campoleoni:2022ebj}. Accordingly, we work in this fixed ADM frame and restrict attention to foliation-preserving diffeomorphisms compatible with the time gauge and the stationary, axisymmetric ansatz. Since possible residual local Carroll boosts are not classified, the solutions should be regarded as gauge-fixed representatives, and individual ADM coefficients should not be interpreted as gauge-invariant observables. Since $\delta\tau_i=-N\,\partial_i\xi^t$, preserving $\tau_i=0$ requires $\partial_i\xi^t=0$ wherever $N\neq0$, unless a compensating boost $\lambda_i=N\,\partial_i\xi^t$ is included; in the stationary axisymmetric sector the residual diffeomorphisms must in addition preserve the ansatz.
\paragraph{Magnetic truncation of ADM: }

The leading-order theory obtained from the small-$c$ expansion of general relativity is the electric Carroll theory, since at this order the action still contains time derivatives of the metric, lapse, and shift. There is also a magnetic theory, which arises as a consistent truncation of the NLO theory. As discussed in Sec.~\ref{staticsec}, the magnetic theory captures many static solutions of general relativity. Let us start from the rescaled action \eqref{rescaledevenADM},
\begin{equation}
\mathcal{L}= \frac{c^{4}}{2\,\kappa}\, \hat N\,\sqrt{\hat \g}\, \left[\hat R\,  + \,c^{-2}\,\left(\hat K_{ij} \hat K^{ij}\,- \,\hat K^2\,\right) \right]\,,\label{rescaledADM}
\end{equation}
and rewrite it in the form
\begin{equation}
\mathcal{L}= \frac{c^{4}}{2\,\kappa}\, \hat N\,\sqrt{\hat \g}\, \left( \hat R  +\, (\hat K_{ij} \chi^{ij}\,- \,\hat K\,\chi)  - \frac{c^{2}}{4}\, (\chi_{ij} \chi^{ij}\,- \,\chi^2)\right)\,.
\end{equation}
Integrating out $\chi_{ij}$ yields an action equivalent to \eqref{rescaledADM}. One then takes the $c\to0$ limit to obtain the magnetic Carroll action in ADM form:
\begin{equation}
\mathcal{L}= \frac{c^{4}}{2\,\kappa}\, N\,\sqrt{\gamma}\, \left(R  +\, \lambda^{ij}\,K_{ij}\right)\,, \label{admmagnetic}
\end{equation}
where $\lambda^{ij} = \chi^{ij}- \g^{ij}\,\chi$. This action is the 3+1 decomposition of the magnetic Carroll action given in \cite{Hansen:2021fxi}. Variation with respect to the ADM Carroll variables yields
\bea 
&\d N:\,&0=R\,, \nn \\
&\d N^i:\,&0=\nabla_j\lambda_i\,^{j}\,,\nn \\
&\d \lambda^{ij}:\, &0= K_{ij}\,,\nn \\
&\d\g^{ij}:\,&0=N G_{ij}-\nabla_{i}\nabla_{j}N+\g_{ij}\nabla^2\, N+\lambda_{k (j}\nabla^kN_{i)}-\frac{1}{2}N^k\nabla_k\lambda_{ij} +\frac{1}{2}\g_{ik}\g_{jl}\del_t\lambda^{kl}\,. \nn 
\eea 
If we set $N^{i}= \lambda^{ij}= 0$, drop the time-derivative terms, and assume the static case, the field equations simplify to
\bea 
\d N:\,&0=&R\\
\d\g^{ij}:\,&0=&N G_{ij}-\nabla_{i}\nabla_{j}N+\g_{ij}\nabla^2 N\,. 
\eea 
Some nontrivial background solutions of magnetic Carroll gravity were obtained in \cite{Hansen:2021fxi,deBoer:2023fnj,Chen:2024how,Kolar:2025ebv}. Using the maps \eqref{map1} and \eqref{map2} again, one can easily show that the action \eqref{admmagnetic} is equivalent to \eqref{covmag}.

\section{NNLO action} \label{NNLOaction}

\paragraph{Compact form of NLO action:} Let us first comment on the NLO action \eqref{NLO}. Up to a boundary term, this action can be written in the compact form
\bea
\mathcal{L}_{NLO}= \frac{\sqrt{\gamma}}{2\,\kappa} \left( N\, R\,   + M\,\overset{(0)}{E}_0 + A^i\,\overset{(0)}{E}_i- \frac{\beta^{ij}}{2}\,\, \overset{(0)}{E}_{ij}\,\right)\,, \label{NLOComp}
\eea
where $\overset{(0)}{E}_0$, $\overset{(0)}{E}_i$, and $\overset{(0)}{E}_{ij}$ are the field equations corresponding to the NLO fields ($M, A^i, \beta^{ij}$):
\bea
\overset{(0)}{E}_0&=&\left(K_{ij}\, K^{ij} - K^2\right)\,\,, \nn \\ 
\overset{(0)}{E}_i\, &=&2\, \nabla^{j}\left( K_{ij} - \gamma_{ij}\, K\right)\,,\nn \\
\overset{(0)}{E}_{ij}&=&N\left(-2K_{i}{}^{k}K_{jk}+K\,K_{ij}+\frac12\left(3 K_{kl}K^{kl}-K^2\right)\gamma_{ij}\right)
-2K_{k(i}\nabla_{j)}N^k\, - N^k\nabla_k K_{ij} \, \nn \\
&&+2\gamma_{ij}K_{kl}\nabla^l N^k
+\,\partial_t K_{ij}
- \,\gamma_{ij}\gamma^{kl}\partial_t K_{kl}
+\,\gamma_{ij}N^k\nabla_k K
\eea
Note that these field equations are also the field equations for the LO action \eqref{LO}. This compact form makes it clear that, when the leading-order field equations ($\overset{(0)}{E}_0 = 0, \overset{(0)}{E}_i = 0, \overset{(0)}{E}_{ij} = 0$) are satisfied, the NLO Lagrangian reduces on shell to the spatial-curvature term alone. This is the basic mechanism behind the magnetic truncation of NLO Carroll gravity discussed in Appendix~\ref{covCar}.


\paragraph{NNLO action:} 

We now turn to the NNLO action. Within the ADM formalism, the NNLO action is obtained straightforwardly by using the same expansion ans\"atze as in \eqref{expand}. The result is
\begin{equation}
\begin{aligned}
\mathcal{L}_{\rm NNLO}=&\sqrt{\gamma}N\Bigg\{\Big(\nabla_j\nabla_i- \gamma_{ij}\,\nabla^2 - G_{ij} \,\Big)\beta^{ij}
\\[4pt]
&+\;\frac{M}{N}\Big[ R
+2\,K^{ij}\theta_{ij}
+\frac12\,(K_{ij}K^{ij}-K^2)\,\beta
\Big] + \frac{P}{N}
(K_{ij}K^{ij}-K^2)
\\[4pt]
&+\,
2\,F^{ij}K_{ij}-2\,F\,K
-2\,K_{i}{}^{k}K^{ij}\epsilon_{jk}
+2\,K\,K^{ij}\epsilon_{ij}
+\frac12\,(K_{ij}K^{ij}-K^2)\,\epsilon
\\
&
+\,(L_{ij}L^{ij}-L^2)
+2\,K^{ij}L\,\beta_{ij}
+K^{ij}L_{ij}\,\beta
-K\,L\,\beta
-4\,K^{ij}L_{i}{}^{k}\beta_{jk}
+2\,K\,L^{ij}\beta_{ij}
\\
&
+\;K^{ij}K^{kl}\beta_{ik}\beta_{jl}
-\;K^{ij}K^{kl}\beta_{ij}\beta_{kl}
+2\,K_{i}{}^{k}K^{ij}\beta_{j}{}^{l}\beta_{kl}
-2\,K\,K^{ij}\beta_{j}{}^{l}\beta_{il}
\\
&
-\frac14\,(K_{ij}K^{ij}-K^2)\,\beta_{kl}\beta^{kl}
-\;K_{i}{}^{k}K^{ij}\beta_{jk}\,\beta
+\;K\,K^{ij}\beta_{ij}\,\beta
+\frac18\,(K_{ij}K^{ij}-K^2)\,\beta^2
\Bigg\}.
\end{aligned}
\end{equation}
Here $\epsilon=\epsilon^i{}_i$ and $\beta^2=\beta^i{}_i\beta^j{}_j$. For stationary configurations, the terms quadratic in $K_{ij}$ can be effectively omitted, since $N^i$ plays a Lagrange-multiplier-like role in the NNLO action. The resulting stationary effective NNLO action is
\bea
\mathcal{L}_{\rm NNLOe}&=& \scalemath{0.9}{\displaystyle{\sqrt{\gamma}N\Bigg\{\Big(\nabla_j\nabla_i- \gamma_{ij}\,\nabla^2 - G_{ij} \,\Big)\beta^{ij} +\;\frac{M}{N}\Big[ R
+2\,K^{ij}\theta_{ij}
\Big] +\;\Big[
2\,F^{ij}(K_{ij}- \gamma_{ij}\,K)}} \, \nn 
\\
&+&\scalemath{0.9}{\,L_{ij}L^{ij}-L^2
+2\,K^{ij}L\,\beta_{ij}
+K^{ij}L_{ij}\,\beta
-K\,L\,\beta
-4\,K^{ij}L_{i}{}^{k}\beta_{jk}
+2\,K\,L^{ij}\beta_{ij}\Big]
\Bigg\}}.
\eea 
The corresponding field equations are given by the reduced equations in \eqref{subNNLO}. Hence, this action may be used as an effective action for stationary backgrounds up to NNLO in the expansion of the parameters appearing in the metric, such as $m$, $G$, $J$, and $Q$.


\section{Field equations} \label{eomfull}
In this appendix, we collect the field equations and their expansion in powers of small $c$. We begin with the relativistic field equations. Varying the action \eqref{preADM} with respect to the ADM fields yields
\bea
\delta\, S_{ADM}= \frac{c^4}{16\pi G}\, \int\, \sqrt{\hat \gamma}\, d^{d+1}\,x\,\left(  \delta \hat N\, E_{0} + \delta\, \hat N^i\, E_i + \delta\, \hat \gamma^{ij}\, E_{ij}\right)\,, \nn
\eea
where the field equations are explicitly given by
\bea
E_0&:& \left( \hat R - \hat K_{ij}\, \hat K^{ij} +  \, \hat K^2 \,\, \right)\,\,,\nn \\ 
E_i\, &:& 2\,\hat \nabla^{j}\left( \hat K_{ij} - \hat \gamma_{ij}\, \hat K\right)\,,\nn \\
E_{ij} &:& \hat N\,\hat G_{ij}\,  -\hat \nabla_{i}\,\hat \nabla_{j}\hat N +\hat \g_{ij}\, \hat \nabla^2\, \hat N -2\, \hat K_{k(i}\hat \nabla_{j)}\hat N^k\, - \hat N^k \hat \nabla_k\, \hat K_{ij}\, \nn \\
&&+  \hat N\left(-2\, \hat K_{i}{}^{k}\, \hat K_{jk}\, +\, \hat K\,\hat K_{ij}\, +\, \frac12\left(3\, \hat K_{kl}\, \hat K^{kl}-\hat K^2\right)\, \hat \gamma_{ij}\right) \, \nn \\
&&+2\, \hat \gamma_{ij}\, \hat K_{kl} \hat \nabla^l \hat N^k
+c^{-1}\,\partial_t \hat K_{ij}
- c^{-1}\,\hat \gamma_{ij} \hat \gamma^{kl}\partial_t \hat K_{kl}
+\,\hat \gamma_{ij}\hat N^k \hat \nabla_k \hat K\,. 
\label{relADMEOM}
\eea
There are several equivalent ways to present the field equations of the ADM action, especially in relation to the constraint and evolution equations in the Hamiltonian formulation. In this work, however, we adopt the standard Lagrangian form of the equations of motion. It would be interesting to derive the NNLO Hamiltonian form of the action and to construct the corresponding constraint algebra for the extended Carrollian symmetries; see the magnetic case in \cite{Perez:2021abf,Perez:2022jpr}.


\subsection{LO, \texorpdfstring{NLO\,}{NLO } and NNLO field equations} \label{eomexpn}
The field equations can be obtained either by expanding the full relativistic equations or by varying the actions with respect to the expanded fields. We start by expanding the equations of motion order by order. The field equations expand as
\bea
E_0 &\to& \overset{(0)}{E}_0 + c^2\, \overset{(2)}{E}_0 + c^4 \, \overset{(4)}{E}_0\,, \nn \\
E_i &\to& \overset{(0)}{E}_i + c^2\, \overset{(2)}{E}_i + c^4 \, \overset{(4)}{E}_i\,, \nn \\
E_{ij} &\to& \overset{(0)}{E}_{ij}+ c^2\, \overset{(2)}{E}_{ij} + c^4 \, \overset{(4)}{E}_{ij}\,, \nn
\eea 
where the explicit forms of the equations are
\bea
\overset{(0)}{E}_0&:& 
K^2 - K_{ij}K^{ij} \nn \\
\overset{(2)}{E}_0&:& R
-2K^{ij}\th_{ij}\nn \\
\overset{(4)}{E}_0&:&-2\left(F^{ij}K_{ij}-FK\right)
-\left(L_{ij}L^{ij}-L^2\right)
-R^{ij}\beta_{ij}
+\nabla_j\nabla_i\beta^{ij}
-\nabla_j\nabla^j\beta\, \nn \\
&&
+2\left(K_i{}^{k}K^{ij}-K K^{jk}\right)\epsilon_{jk}
+2\left(2K^{ij}L_i{}^{k}-L K^{jk}-K L^{jk}\right)\beta_{jk}\,\nn  \\ 
&&
+\left(
K^{ij}K^{kl}\beta_{ij}\beta_{kl} -K^{ij}K^{kl}\beta_{ik}\beta_{jl}
\right)
-2\left(K_i{}^{k}K^{ij}-K K^{jk}\right)\beta_j{}^{l}\beta_{kl}\, \nn \\ 
\overset{(0)}{E}_i &:&\quad 2\left(\nabla_j K_i{}^{j}-\nabla_i K\right) \nn \\
\overset{(2)}{E}_i&:& \quad 2\beta^{jk}\left(\nabla_i K_{jk}-\nabla_k K_{ij}\right)
+K^{jk}\nabla_i\beta_{jk}
+2\left(\nabla_j L_i{}^{j}-\nabla_i L\right)
+K_i{}^{j}\nabla_j\beta
-2K_i{}^{j}\nabla_k\beta_j{}^{k}\, \nn \\
\overset{(4)}{E}_i&:& \quad 2\left(\nabla_j F_i{}^{j}-\nabla_i F\right)
+2\epsilon^{jk}\left(\nabla_i K_{jk}-\nabla_k K_{ij}\right)
-2\beta_j{}^{l}\beta^{jk}\nabla_i K_{kl}
+2\beta_j{}^{l}\beta^{jk}\nabla_l K_{ik}\, \nn \\
&&+K^{jk}\nabla_i\epsilon_{jk}
+L^{jk}\nabla_i\beta_{jk}
+2\beta^{jk}\left(\nabla_i L_{jk}-\nabla_k L_{ij}\right)
+K_i{}^{j}\nabla_j\epsilon
+L_i{}^{j}\nabla_j\beta\, \nn \\
&&-2K_i{}^{j}\nabla_k\epsilon_j{}^{k}
-2L_i{}^{j}\nabla_k\beta_j{}^{k}
-2K^{jk}\beta_j{}^{l}\nabla_i\beta_{kl}
-K_i{}^{j}\beta^{kl}\nabla_j\beta_{kl}
-K_i{}^{j}\beta_j{}^{k}\nabla_k\beta\, \nn \\
&&+2K_i{}^{j}\beta^{kl}\nabla_l\beta_{jk}
+2K_i{}^{j}\beta_j{}^{k}\nabla_l\beta_k{}^{l}\,.\nn \\ 
\overset{(0)}{E}_{ij}&:&N\left(-2K_{i}{}^{k}K_{jk}+K\,K_{ij}+\frac12\left(3K_{kl}K^{kl}-K^2\right)\gamma_{ij}\right)
-2K_{k(i}\nabla_{j)}N^k \nn \\
&&-N^k\nabla_k K_{ij}
+2\gamma_{ij}K_{kl}\nabla^l N^k
+\partial_t K_{ij}
-\gamma_{ij}\gamma^{kl}\partial_t K_{kl}
+\gamma_{ij}N^k\nabla_k K \nn \\
\overset{(2)}{E}_{ij}&:& \partial_t L_{ij} +M\left(
-2K_{i}{}^{k}K_{jk}
+K\,K_{ij}
+\frac12\left(3K_{kl}K^{kl}-K^2\right)\gamma_{ij}
\right)
\nn \\[4pt]
&&+N\left(
R_{ij}
+K\,L_{ij}
+L\,K_{ij}
-4K_{(i}{}^{k}L_{j)k}
+\frac12\left(3K_{kl}K^{kl}-K^2\right)\beta_{ij}
+2K_{i}{}^{k}K_{j}{}^{l}\beta_{kl}
\right.
\nn \\[4pt]
&&\left.
\quad-K_{ij}K^{kl}\beta_{kl} +\left(3K^{kl}L_{kl}-KL-\frac12 R\right)\gamma_{ij}
-3K_{k}{}^{m}K^{kl}\beta_{lm}\,\gamma_{ij}
+K\,K^{lm}\beta_{lm}\,\gamma_{ij}
\right)
\nn \\[4pt]
&&-\gamma^{kl}\partial_t K_{kl}\,\beta_{ij}
-\gamma^{kl}\partial_t L_{kl}\,\gamma_{ij}
+\beta^{kl}\partial_t K_{kl}\,\gamma_{ij}
\nn \\[4pt]
&&-2K_{k(i}\nabla_{j)}A^k
-2L_{k(i}\nabla_{j)}N^k
-\nabla_i\nabla_j N
-A^k\nabla_k K_{ij}
-N^k\nabla_k L_{ij}
\nn \\[4pt]
&&+\beta_{ij}\left(
N^k\nabla_k K
+2K_{kl}\nabla^l N^k
\right)
\nn \\[4pt]
&&+\gamma_{ij}\left(
A^k\nabla_k K
+N^k\nabla_k L
+\nabla_k\nabla^k N
+2K_{kl}\nabla^l A^k
+2L_{kl}\nabla^l N^k
\right)
\nn \\[4pt]
&&-\gamma_{ij}\left(
N^k\beta^{lm}\nabla_k K_{lm}
+2K_{k}{}^{m}\beta_{lm}\nabla^l N^k
\right)\,\nn \\
\overset{(4)}{E}_{ij}&:& \quad
\mathcal{E}^{(0)}_{ij}
+\mathcal{E}^{(1)}_{ij}
+\mathcal{E}^{(2)}_{ij}
+\mathcal{E}^{(\nabla)}_{ij}\,,
\eea 
where
\[
\begin{aligned}
\mathcal{E}^{(0)}_{ij}=&
\partial_t F_{ij}
-\gamma^{kl}\partial_t K_{kl}\,\epsilon_{ij}
\\[2pt]
&+M\left(
R_{ij}
+K\,L_{ij}
+L\,K_{ij}
-4K_{k(i}L_{j)}{}^{k}
\right)
+P\left(
-2K_{i}{}^{k}K_{jk}
+K\,K_{ij}
\right)
\\[2pt]
&+N\left(
F\,K_{ij}
+K\,F_{ij}
-4F_{k(i}K_{j)}{}^{k}
+\frac32 K_{kl}K^{kl}\,\epsilon_{ij}
-\frac12 K^2\,\epsilon_{ij}
+2K_{i}{}^{k}K_{j}{}^{l}\epsilon_{kl}
-K_{ij}K^{kl}\epsilon_{kl}
\right.
\\[2pt]
&\qquad\left.
-2L_{i}{}^{k}L_{jk}
+L\,L_{ij}
\right)
\\[4pt]
&+\gamma_{ij}\left(
-\gamma^{kl}\partial_t F_{kl}
+\epsilon^{kl}\partial_t K_{kl}
+3M\,K^{kl}L_{kl}
-M\,K\,L
+\frac32 P\,K_{kl}K^{kl}
-\frac12 P\,K^2
-\frac12 M\,R
\right.
\\[2pt]
&\qquad\left.
+3N\,F^{kl}K_{kl}
-N\,F\,K
-3N\,K_{k}{}^{m}K^{kl}\epsilon_{lm}
+N\,K\,K^{lm}\epsilon_{lm}
+\frac32 N\,L_{kl}L^{kl}
-\frac12 N\,L^2
\right),
\end{aligned}
\]
\[
\begin{aligned}
\mathcal{E}^{(1)}_{ij}=&
-\gamma^{kl}\partial_t L_{kl}\,\beta_{ij}
+\left(
\frac32 M\,K_{kl}K^{kl}
-\frac12 M\,K^2
+3N\,K^{kl}L_{kl}
-N\,K\,L
-\frac12 N\,R
\right)\beta_{ij}
\\[2pt]
&+M\left(
2K_{i}{}^{k}K_{j}{}^{l}
-K_{ij}K^{kl}
\right)\beta_{kl} +N\left(
4K_{(i}{}^{k}L_{j)}{}^{l}
-K_{ij}L^{kl}
-L_{ij}K^{kl}
\right)\beta_{kl}
\\[2pt]
&+\gamma_{ij}\left(
\beta^{kl}\partial_t L_{kl}
+N\,L\,K^{kl}\beta_{kl}
+\frac12 N\,R^{kl}\beta_{kl}
-3M\,K_{k}{}^{m}K^{kl}\beta_{lm}
+M\,K\,K^{lm}\beta_{lm}
\right.
\\[2pt]
&\qquad\left.
-6N\,K^{kl}L_{k}{}^{m}\beta_{lm}
+N\,K\,L^{lm}\beta_{lm}
\right),\, \nn \\
\mathcal{E}^{(2)}_{ij}=&
\;\beta_{ij}\,\beta^{kl}\partial_t K_{kl}
+N\,\beta_{ij}\left(
-3K_{k}{}^{m}K^{kl}\beta_{lm}
+K\,K^{lm}\beta_{lm}
\right)\, \nn \\
&\qquad+N\left(
-2K_{i}{}^{k}K_{j}{}^{l}
+K_{ij}K^{kl}
\right)\beta_{k}{}^{m}\beta_{lm}
\\[2pt]
&+\gamma_{ij}\left(
-\beta_{k}{}^{m}\beta^{ln}\,\partial_t K_{mn}
+\frac32 N\,K^{kl}K^{mn}\beta_{km}\beta_{ln}
-\frac12 N\,K^{kl}K^{mn}\beta_{kl}\beta_{mn}
\right.
\\[2pt]
&\qquad\left.
+3N\,K_{k}{}^{m}K^{kl}\beta_{l}{}^{n}\beta_{mn}
-N\,K\,K^{lm}\beta_{l}{}^{n}\beta_{mn}
\right),
\end{aligned}
\]
\[
\begin{aligned}
\mathcal{E}^{(\nabla)}_{ij}=&
-2L_{k(i}\nabla_{j)}A^{k}
-2K_{k(i}\nabla_{j)}Z^{k}
-2F_{k(i}\nabla_{j)}N^{k}
-\nabla_i\nabla_j M
\\
&-N\,\nabla_i\nabla_j\beta
+N\,\nabla_k\nabla_{(i}\beta_{j)}{}^{k}
-\frac12 N\,\nabla_k\nabla^k\beta_{ij}
\\
&+\gamma_{ij}\left(
\nabla_k\nabla^k M
-\frac12 N\,\nabla_k\nabla_l\beta^{kl}
+\frac12 N\,\nabla_k\nabla^k\beta
\right)
+\beta_{ij}\nabla_k\nabla^k N
-\gamma_{ij}\beta_{kl}\nabla^l\nabla^k N
\\
&-N^{k}\nabla_k F_{ij}
+\gamma_{ij}N^{k}\nabla_k F
-Z^{k}\nabla_k K_{ij}
-N^{k}\beta_{ij}\beta^{lm}\nabla_k K_{lm}
-N^{k}\epsilon^{lm}\gamma_{ij}\nabla_k K_{lm}
\\
&-A^{k}\beta^{lm}\gamma_{ij}\nabla_k K_{lm}
+N^{k}\epsilon_{ij}\nabla_k K
+A^{k}\beta_{ij}\nabla_k K
+Z^{k}\gamma_{ij}\nabla_k K
\\
&+N^{k}\beta_{l}{}^{m}\beta^{ln}\gamma_{ij}\nabla_k K_{mn}
-A^{k}\nabla_k L_{ij}
-N^{k}\beta^{lm}\gamma_{ij}\nabla_k L_{lm}
+N^{k}\beta_{ij}\nabla_k L
+A^{k}\gamma_{ij}\nabla_k L
\\
&+2K_{kl}\beta_{ij}\nabla^{l}A^{k}
+2L_{kl}\gamma_{ij}\nabla^{l}A^{k}
-2K_{k}{}^{m}\beta_{lm}\gamma_{ij}\nabla^{l}A^{k}
+2K_{kl}\gamma_{ij}\nabla^{l}Z^{k}
\\
&+2K_{kl}\epsilon_{ij}\nabla^{l}N^{k}
+2L_{kl}\beta_{ij}\nabla^{l}N^{k}
-2K_{k}{}^{m}\beta_{ij}\beta_{lm}\nabla^{l}N^{k}
+2F_{kl}\gamma_{ij}\nabla^{l}N^{k}
\\
&-2K_{k}{}^{m}\epsilon_{lm}\gamma_{ij}\nabla^{l}N^{k}
-2L_{k}{}^{m}\beta_{lm}\gamma_{ij}\nabla^{l}N^{k}
+2K_{k}{}^{m}\beta_{l}{}^{n}\beta_{mn}\gamma_{ij}\nabla^{l}N^{k}
-\beta_{kl}\gamma_{ij}\nabla^{l}\nabla^{k}N.
\end{aligned}
\]

\subsection{Stationary reduction and reduced field equations} \label{eomreduce}
 
The stationary reduction of the field equations is obtained by imposing time independence of all fields and setting the leading-order shift vector to zero, $N^i=0$. The NLO field equations then reduce to
\begin{align}
\delta N: \quad & R = 0\,,  \nonumber \\
\delta N^i: \quad & 2\,\nabla^j(L_{ij} - \gamma_{ij} L) = 0 \,,  \nonumber \\
\delta \gamma^{ij}: \quad & N G_{ij} - \nabla_i \nabla_j N + \gamma_{ij} \nabla^2 N = 0\,, \label{subNLO}
\end{align}
where the $L$ tensor is understood to be evaluated in this limit ($L \to L_{|_{N_i=0}}$, etc.). This system strongly constrains the solution space. The only subleading field that appears explicitly at this order is the vector field $A_i$, which arises from the NLO part of the shift vector. The $\delta N_i$ equation is therefore the main feature that distinguishes the full NLO theory from the magnetic truncation, in which $N^i=0$ and no NLO vector field is present. In this sense, the full NLO theory already exhibits a richer structure by allowing rotational degrees of freedom. Furthermore, the remaining NNLO field equations are simplified in the stationary limit as follows:
\begin{align}
\delta N:\, & L^2 - L_{ij} L^{ij} - R^{ij} \beta_{ij} + \nabla_j \nabla_i \beta^{ij} - \nabla^2 \beta = 0 \,, \nonumber \\[4pt]
\delta N^i:\,  & 2\, \nabla^j(F_{ij} - \gamma_{ij} F) + 2 \beta_{jk}(\nabla_i L^{jk} - \nabla^j L_i{}^k) + L_{jk} \nabla_i \beta^{jk} + L_{ij} \nabla^j \beta - 2 L_{ik} \nabla_j \beta^{kj} = 0\,,   \nonumber \\[4pt]
\delta \gamma^{ij}:\,   &M\,G_{ij}
+\left(\g_{ij}\nabla^2-\nabla_i \nabla_j\right) M
+\left(2L_{kl}\nabla^l-2 L_{k(j}\nabla_{i)} \right)A^k+A^k\nabla_k\left(
L\g_{ij}- L_{ij} \right)\nn
\\[4pt]
&+\,N\Bigg(
-2L_i{}^k L_{jk}
+L\,L_{ij}
-\frac12 R\,\beta_{ij}
-\frac12 \nabla_i\nabla_j \beta
+\nabla_l\nabla_{(i} \beta_{j)}{}^l
-\frac12 \nabla^2 \beta_{ij}
\Bigg)\nn
\\[4pt]
&+\,\gamma_{ij}\frac{N}{2}\left(3 L_{kl}L^{kl}
- L^2
+ R^{kl}\beta_{kl}
- \nabla_m\nabla_l \beta^{lm}
+ \nabla^2 \beta  \right) \nn \\
& +\,\left(\beta_{ij}\nabla^2-\g_{ij}\b^{k l}\nabla_k\nabla_l\right) N 
=0 \, ,  \label{subNNLO}
\end{align}
where, in addition to $L$, the $F$ tensor is understood to be evaluated in this limit ($F \to F_{|_{N_i=0}}$, etc.). These field equations are our starting point for discussing various nontrivial background solutions in the main text.

\subsection{Further reduction of field equations} \label{eomfurtreduce}
We can simplify the field equations further by imposing $A_i = 0$. This reduction is useful for constructing the weak-branch NNLO solutions of the Kerr, Hartle--Thorne, and mixed types. In this case, the reduced equations \eqref{subNNLO} take the following linear form:
\begin{align}
\delta N:\, & \nabla_j \nabla_i \beta^{ij} - \nabla^2 \beta  - R^{ij} \beta_{ij} = 0 \,, \nonumber \\
\delta N^i:\,  & 2\, \nabla^j(F_{ij} - \gamma_{ij} F) = 0\,,   \nonumber \\
\delta \gamma^{ij}:\,   &M\,G_{ij}+\left(\g_{ij}\nabla^2
-\nabla_i\nabla_j\right) M
+\beta_{ij}\nabla^2 N-\g_{ij}\b_{kl}\nabla^k\nabla^lN \nn
\\[4pt]
&\quad+\,N\Bigg(
-\frac12 R\,\beta_{ij}
-\frac12 \nabla_i\nabla_j \beta
+\nabla_k\nabla_{(i} \beta_{j)}{}^{k}
-\frac12 \nabla^2 \beta_{ij}
\Bigg) \nn
\\[4pt]
&\quad+\,\frac{\gamma_{ij}N}{2}\Bigg[
 \,R^{kl}\beta_{kl}
- \,\nabla_l\nabla_k \beta^{kl}
+ \,\nabla^2 \beta \Bigg]
=0\,.\label{subsubNNLO}
\end{align}

\section{Lense--Thirring metric} \label{LT}
A spinning body induces frame dragging in spacetime. For example, the orbit of a nearby test particle precesses around the rotating source. This effect is the gravitational analogue of magnetic induction and was first predicted by Lense and Thirring in 1918 \cite{Lense:1918}. The well-known Lense--Thirring metric is
\bea
ds^2 = -f\,c^2 dt^2 + \frac{dr^2}{f} + 2\,c\,a\,(f - 1)\sin^2\theta\,dt\,d\phi 
+ r^2\left(\sin^2\theta\,d\phi^2 + d\theta^2\right),
\eea 
where $f=1-\frac{2Gm}{rc^2}$ and, in conventional units, $a=J/(mc)$. This metric solves the vacuum Einstein equations to leading order in the rotation parameter $a$ and describes the spacetime outside a slowly rotating body. It also arises as the slow-rotation limit of both the Hartle--Thorne and Kerr metrics. In the strong-gravity scaling considered here, one may take
\[J \to c^3\,J\,, \qquad m \to c\, m\,,\qquad G\to c\, G\]
and then read off the corresponding small-$c$ expansion of the ADM fields as given in Section~\ref{LTsol}.

\section{Hartle--Thorne metric} \label{HT}
The Hartle--Thorne metric is an approximate solution of the vacuum Einstein equations describing the exterior field of a slowly rotating, weakly deformed compact object. As discussed in Section~\ref{HTweak}, its weak-field expansion gives rise to an exact solution of the NNLO theory. Let us first present the relativistic Hartle--Thorne line element in the form \cite{Hartle:1967he,Hartle:1968si}
\begin{align}
ds^{2} &=
-\left(1-\frac{2m}{r}\right)\!\Bigg[1+2k_{1}P_{2}(\cos\theta)
+2\left(1-\frac{2m}{r}\right)^{-1}\frac{J^{2}}{r^{4}}\,(2\cos^{2}\theta-1)\Bigg]\,dt^{2} \notag\\
&\quad
+\left(1-\frac{2m}{r}\right)^{-1}\!\Bigg[1-2k_{2}P_{2}(\cos\theta)
-2\left(1-\frac{2m}{r}\right)^{-1}\frac{J^{2}}{r^{4}}\Bigg]\,dr^{2} \notag\\
&\quad
+r^{2}\,[1-2k_{3}P_{2}(\cos\theta)]\,(d\theta^{2}+\sin^{2}\theta\,d\phi^{2})
-\frac{4J}{r}\sin^{2}\theta\,dt\,d\phi\,,
\end{align}
where
\begin{align}
k_{1} &= \frac{J^{2}}{mr^{3}}\!\left(1+\frac{m}{r}\right)
+\frac{5}{8}\,\frac{Q-\tfrac{J^{2}}{m}}{m^{3}}\,
Q^{2}_{2}\!\left(\frac{r}{m}-1\right)\,, \nn \\
k_{2} &= k_{1}-\frac{6J^{2}}{r^{4}}\,,  \nn \\
k_{3} &= k_{1}+\frac{J^{2}}{r^{4}}
+\frac{5}{4}\,\frac{Q-\tfrac{J^{2}}{m}}{m^{2}\sqrt{r^{2}-2mr}}\,
Q^{1}_{2}\!\left(\frac{r}{m}-1\right)\,, \nn \\
P_{2}(\cos\theta) &= \tfrac12\,(3\cos^{2}\theta-1)\,, \nn \\
Q^{1}_{2}(x) &= (x^{2}-1)^{1/2}\!\left[\frac{3x}{2}\ln\!\frac{x+1}{x-1}
-\frac{3x^{2}-2}{x^{2}-1}\right]\,, \nn \\
Q^{2}_{2}(x) &= (x^{2}-1)\!\left[\frac{3}{2}\ln\!\frac{x+1}{x-1}
-\frac{3x^{3}-5x}{(x^{2}-1)^{2}}\right].
\end{align}
To restore dimensions explicitly, we substitute
\[
m \to \frac{Gm}{c^2},\qquad J \to \frac{GJ}{c^3},\qquad Q \to \frac{GQ}{c^2}.
\]
The line element then becomes
\begin{align}
ds^{2} &=
-\left(1-\frac{2mG}{c^2 r}\right)\!\Bigg[1+2k_{1}P_{2}(\cos\theta)
+2\left(1-\frac{2mG}{c^2r}\right)^{-1}\frac{G^2J^{2}}{c^6r^{4}}\,(2\cos^{2}\theta-1)\Bigg]\,c^2 dt^{2} \notag\\
&\quad
+\left(1-\frac{2m G}{c^2 r}\right)^{-1}\!\Bigg[1-2k_{2}P_{2}(\cos\theta)
-2\left(1-\frac{2mG}{c^2r}\right)^{-1}\frac{G^2J^{2}}{c^6r^{4}}\Bigg]\,dr^{2} \notag\\
&\quad
+r^{2}\,[1-2k_{3}P_{2}(\cos\theta)]\,(d\theta^{2}+\sin^{2}\theta\,d\phi^{2})
-\frac{4GJ}{c^2r}\sin^{2}\theta\, dt\,d\phi 
\end{align}
where
\begin{align}
k_{1} &= \frac{G J^{2}}{c^4 m r^{3}}\!\left(1+\frac{Gm}{c^2r}\right)
+\frac{5c^2}{8}\,\frac{GQc^2-\tfrac{GJ^{2}}{m}}{G^3m^{3}}\,
Q^{2}_{2}\!\left(\frac{c^2r}{Gm}-1\right)\,, \nn \\
k_{2} &= k_{1}-\frac{6G^2J^{2}}{c^6r^{4}}\,,  \nn \\
k_{3} &= k_{1}+\frac{G^2J^{2}}{c^6r^{4}}
+\frac{5c^2}{4}\,\frac{GQ-\tfrac{GJ^{2}}{c^2m}}{G^2m^{2}\sqrt{r^{2}-\tfrac{2Gmr}{c^2}}}\,
Q^{1}_{2}\!\left(\frac{c^2r}{Gm}-1\right)\,, \nn \\
P_{2}(\cos\theta) &= \tfrac12\,(3\cos^{2}\theta-1)\,, \nn \\
Q^{1}_{2}(x) &= (x^{2}-1)^{1/2}\!\left[\frac{3x}{2}\ln\!\frac{x+1}{x-1}
-\frac{3x^{2}-2}{x^{2}-1}\right]\,, \nn \\
Q^{2}_{2}(x) &= (x^{2}-1)\!\left[\frac{3}{2}\ln\!\frac{x+1}{x-1}
-\frac{3x^{3}-5x}{(x^{2}-1)^{2}}\right].
\end{align}
The Hartle--Thorne metric admits two distinct scaling regimes in the present context.

\begin{enumerate}
\item \textbf{Strong-gravity branch.} This branch corresponds to the scaling
\[
J\to c^3J,\qquad m\to c\, m,\qquad G\to c\, G.
\]
In this regime, one may focus on the black-hole-type relation $Q=J^2/(Gm)$. The metric then takes the form
\begin{align}
ds^{2} &=
-\left(1-\frac{2mG}{ r}\right)\!\Bigg[1+2k_{1}P_{2}(\cos\theta)
+2\left(1-\frac{2mG}{r}\right)^{-1}\frac{c^2 G^2J^{2}}{r^{4}}\,(2\cos^{2}\theta-1)\Bigg]\,c^2 dt^{2} \notag\\
&\quad
+\left(1-\frac{2m G}{ r}\right)^{-1}\!\Bigg[1-2k_{2}P_{2}(\cos\theta)
-2\left(1-\frac{2mG}{r}\right)^{-1}\frac{c^2G^2J^{2}}{r^{4}}\Bigg]\,dr^{2} \notag\\
&\quad
+r^{2}\,[1-2k_{3}P_{2}(\cos\theta)]\,(d\theta^{2}+\sin^{2}\theta\,d\phi^{2})
-\frac{4c^2GJ}{r}\sin^{2}\theta\, dt\,d\phi
\end{align}
where
\begin{align}
k_{1} &= \frac{c^2 G J^{2}}{mr^{3}}\!\left(1+\frac{mG}{r}\right)\,, \nn \\
k_{2} &= k_{1}-\frac{6c^2G^2J^{2}}{ r^{4}}\,, \nn \\
k_{3} &= k_{1}+\frac{c^2G^2J^{2}}{r^{4}}\,. 
\end{align}
Therefore, using \eqref{expand}, at orders $c^0$ and $c^2$ we obtain the following match:
\begin{equation}
\begin{aligned}
N &= \sqrt{1 - \frac{2Gm}{r}}\,, \quad M =- \frac{2GJ^2}{mr^3\,\sqrt{1 - \frac{2Gm}{r}}}  P_2(\cos\theta)\,, \\
\g_{ij}&=\text{\rm diag}\left( \frac{1}{1 - \frac{2Gm}{r} }, r^2, r^2\sin^2{\th} \right)\,, \quad A_{\phi} = -\frac{2GJ}{r} \sin^2\theta\,. 
\end{aligned}
\end{equation}
The $M=\mathcal{O}(J^2)$ term can be read off explicitly from the expansion of the Hartle--Thorne metric. However, as already noted, the NLO field equations with $N^i=0$ leave $M$ undetermined, so it may consistently be set to zero at that order. One might expect the $J^2$ terms to be captured by the NNLO Carroll action, but as discussed in Section~\ref{LTNNLOSol}, this does not lead to a consistent strong-branch NNLO solution. For that reason, in order to capture the strong-gravity $J^2$ and $J^3$ structure consistently, we instead turn to the Kerr metric.

\item \textbf{Weak-field branch.} This branch corresponds to the scaling
\[
J\to c^4J,\qquad m\to c^2m,\qquad G\to c^2G,\qquad Q\to c^2Q,
\]
where the quadrupole moment $Q$ is kept independent of the rotation parameter. The line element is given by
\begin{equation}
\begin{aligned}
\frac{g_{tt}}{c^2}=&-1 +c^2\left(\frac{2G m}{r}-\frac{2GQ P_2(\cos{\th})}{r^3} \right)-c^4\frac{2G^2mQ P_2(\cos{\th})}{r^4} \\
g_{t\phi}=& -c^4\frac{2GJ\sin^2{\th}}{r}\\
g_{rr}=&1+c^2\left( \frac{2G m}{r}-\frac{2GQP_2(\cos{\th})}{r^3} \right)+c^4\frac{2G^2}{r^4}\left( 2m^2 r^2-5mQP_2(\cos{\th}) \right) \\
g_{\th\th}=& r^2-c^2\frac{2 GQP_2(\cos{\th})}{r}-c^4\frac{5 G^2 m r^2 Q P_2(\cos{\th})}{r^4}\\
g_{\phi\phi}=& r^2\sin^2{\th}-c^2\frac{2 GQP_2(\cos{\th})}{r}\sin^2{\th}-c^4\frac{5 G^2 m r^2 Q P_2(\cos{\th})}{r^4}\sin^2{\th}
\end{aligned}
\end{equation}
Then, we read the fields as follows:
\begin{equation}
\begin{aligned}
&N=1,\quad M= -\frac{Gm}{r}+ \frac{GQP_2(\cos{\th})}{r^3}\,,\quad Z_\phi=-\frac{2GJ}{r}\sin^2{\th}\,,\\
&\g_{ij}=\text{\rm diag}\left( 1, r^2, r^2\sin^2{\th} \right)\,,\\
&\b_{rr}=\frac{2Gm}{r}-\frac{2GQ P_2(\cos{\th})}{r^3}\,,\\
&\b_{\th\th}=-\frac{2 GQP_2(\cos{\th})}{r} \,,\quad\b_{\phi\phi}=\b_{\th\th}\,\sin^2{\th} \,.
\end{aligned}
\end{equation}
This corresponds to the solution given in Section~\ref{HTweak}.
\end{enumerate}

\section{Kerr metric} \label{Kerr}
In this appendix, we expand the Kerr metric in order to recover both the Lense--Thirring solution and the $J^2$ contributions appearing in NNLO Carroll gravity. Let us begin with the Kerr metric in Boyer--Lindquist coordinates \cite{Kerr:1963ud}:

\begin{equation}
\begin{aligned}
ds^{2}
= {}&
-\left(1-\frac{2Gm r}{\Sigma c^{2}}\right)c^{2}dt^{2}
-\frac{4Gm a r \sin^{2}\theta}{\Sigma c^{2}}\,c\,dt\,d\phi\, \nn \\
&
+\frac{\Sigma}{\Delta}\,dr^{2}
+\Sigma\,d\theta^{2}
+\left(r^{2}+a^{2}
+\frac{2Gm a^{2} r \sin^{2}\theta}{\Sigma c^{2}}\right)
\sin^{2}\theta\,d\phi^{2},
\end{aligned}
\end{equation}
where
\bea
\Sigma = r^{2}+a^{2}\cos^{2}\theta,
\qquad
\Delta = r^{2}-\frac{2Gm r}{c^{2}}+a^{2}, \qquad a = \frac{J}{m c}.
\eea 

\begin{enumerate}

\item \textbf{Strong-gravity branch.} We impose the scalings
\[
J\to c^3\,J,\qquad m\to c\,m,\qquad G\to c\,G.
\]
The Kerr metric then becomes
\[
\begin{aligned}
ds^{2} =&\;
-c^{2}\left(
1-\frac{2Gm r}{\,r^{2}+c^{2}\dfrac{J^{2}}{m^{2}}\cos^{2}\theta\,}
\right)dt^{2}
-\frac{4c^{2}GJ r\sin^{2}\theta}{\,r^{2}+c^{2}\dfrac{J^{2}}{m^{2}}\cos^{2}\theta\,}\;dt\,d\phi
\\[4pt]
&\;
+\frac{\,r^{2}+c^{2}\dfrac{J^{2}}{m^{2}}\cos^{2}\theta\,}{\,r^{2}-2Gm r+c^{2}\dfrac{J^{2}}{m^{2}}\,}\;dr^{2}
+\left(r^{2}+c^{2}\dfrac{J^{2}}{m^{2}}\cos^{2}\theta\right)d\theta^{2}
\\[4pt]
&\;
+\left[
r^{2}+c^{2}\dfrac{J^{2}}{m^{2}}
+\frac{2c^{2}GJ^{2} r\sin^{2}\theta}{\,m\left(r^{2}+c^{2}\dfrac{J^{2}}{m^{2}}\cos^{2}\theta\right)}
\right]\sin^{2}\theta\,d\phi^{2}
\end{aligned}
\]
Expanding all metric components around $c=0$ through order $c^4$, we obtain
\bea 
g_{tt}
&=&
-c^{2}\left[
\left(1-\frac{2Gm}{r}\right)
+ c^{2}\,
\frac{2GJ^{2}}{m r^{3}}\cos^{2}\theta- c^4\frac{2 G J^4\cos^4{\th}}{m^3 r^5}\right]\,, \nn \\
g_{t\phi}
&=& -\frac{2GJ}{r}\,c^{2}\sin^{2}\theta+c^4\frac{2 G J^3 \sin^2{\th}\cos^2{\th}}{m^2 r^3}\,, \nn \\
g_{rr}
&=&
\frac{1}{1-\frac{2Gm}{r}}
\;+\;
c^{2}\,\frac{J^{2}}{m^{2}}\,
\frac{
\cos^{2}\theta\left(1-\dfrac{2Gm}{r}\right)-1
}{
r^{2}\left(1-\dfrac{2Gm}{r}\right)^{2}
}+c^4\frac{J^4}{m^4 r^2}\frac{r-\cos^2{\th}\left(r-2Gm\right)}{\left(r-2 G m\right)^3}\,, \nn \\
g_{\theta\theta}
&=&
r^{2}
+
c^{2}\frac{J^{2}}{m^{2}}\cos^{2}\theta\,, \nn \\
g_{\phi\phi}
&=&
r^{2}\sin^{2}\theta
+
c^{2}\frac{J^{2}}{m^{2}}\sin^{2}\theta
\left(
1+\frac{2Gm}{r}\sin^{2}\theta
\right)-2 c^4\frac{GJ^4\cos^2{\th}\sin^4{\th}}{m^3 r^3}\,. 
\eea 
Using the ADM expansion \eqref{expand}, one can then read off the corresponding Carrollian fields as
\bea
\gamma_{ij}
&=&\text{\rm diag}
\left(
\frac{1}{\,1-\dfrac{2Gm}{r}\,},
\; r^{2},
\; r^{2}\sin^{2}\theta
\right)\,, \nn \\
\beta_{ij}
&=&
\frac{J^2}{m^2}\,\text{\rm diag}\left(\frac{\left(1-\frac{2 Gm}{r}\right)\cos^2{\th}-1}{\left(r-2Gm\right)^2},
\;
\cos^{2}\theta,
\;
\sin^{2}\theta\!
\left(1+\frac{2Gm}{r}\sin^{2}\theta\right)
\right)\,, \nn \\
\e_{ij}
&=&\text{\rm diag}
\left(\frac{J^4}{m^4 r^2}\frac{r-\cos^2{\th}\left(r-2Gm\right)}{\left(r-2 G m\right)^3},
\;
0,
\;
-2\frac{GJ^4\cos^2{\th}\sin^4{\th}}{m^3 r^3}
\right)\,, \nn \\
A_\phi&=&-\frac{2GJ}{r}\sin^2{\th}\,,\qquad Z_\phi  =\frac{2GJ^3}{m^2r^3}\sin^2{\th}\left(1+\cos^2{\th}+\frac{2G m}{r}\sin^2{\th} \right)\,, \nn \\
N&=&\sqrt{1-\frac{2Gm}{r}}, \qquad
M
=
\frac{1}{\sqrt{1-\frac{2Gm}{r}}}
\left(
\frac{GJ^{2}}{m r^{3}}\cos^{2}\theta
+
\frac{2G^{2}J^{2}}{r^{4}}\sin^{2}\theta
\right)\,, \nn \\
P&=& \scalemath{0.85}{ G J^4\frac{12G^3 m^3\sin^4{\th}+20 G^2 m^2 r\cos^2{\th}\sin^2{\th}+ G m r^2\left(11\cos^4{\th}-4\cos^2{\th}-4 \right) -2 r^3\cos^4{\th}  }{2 m^3 r^7\left(r-2 G m\right)\sqrt{1-\frac{2G m}{r}}}}\,. \nn 
\eea 
This expansion gives precisely the vacuum NNLO Carroll solution shown in \eqref{NNLOLT}. In this scaling regime, the small-$c$ expansion is naturally tied to the slow-rotation expansion in powers of $J$. From this perspective, the Carrollian expansion also provides a systematic way to organize rotating metrics order by order in the rotation parameter. More generally, it provides a finite-order framework for organizing exterior rotational deformations motivated by relativistic compact-object geometries.

\item \textbf{Weak-field branch.} We now impose the scalings
\[
J\to c^4\,J,\qquad m\to c^2\,m,\qquad G\to c^2\,G.
\]
This corresponds to the weak-field branch of the Kerr expansion. The metric then takes the form
\bea 
\frac{g_{tt}}{c^2}&=&-1 +  \frac{2Gm c^2}{r} \,, \nn \\
g_{t\phi}&=& -c^4\frac{2GJ\sin^2{\th}}{r}\,, \nn \\
g_{rr}&=&1+c^2\left( \frac{2G m}{r}\right)- \frac{c^2\, J^2}{m^2\,r^2} \sin^2{\th}+c^4\frac{2G^2}{r^4}\left( 2m^2 r^2 \right)  \,, \nn \\
g_{\th\th}&=& r^2 + \frac{c^2\, J^2}{m^2} \cos^2{\th}\,, \nn \\
g_{\phi\phi}&=& r^2\sin^2{\th}+ \frac{c^2\, J^2}{m^2} \sin^2{\th} \,.
\eea 
From this expansion, one reads off the ADM variables as
\begin{equation}
\begin{aligned}
&N=1,\quad M= -\frac{Gm}{r}\,,\quad Z_\phi=-\frac{2GJ}{r}\sin^2{\th}\,,\\
&\g_{ij}=\text{\rm diag}\left(1, r^2, r^2\sin^2{\th}\right)\,,\\
&\b_{rr}=\frac{2Gm}{r} - \frac{J^2}{m^2\,r^2} \sin^2{\th}\,,\\
&\b_{\th\th}= \frac{J^2}{m^2} \cos^2{\th}\,,\quad \b_{\phi\phi}=\frac{J^2}{m^2} \sin^2{\th}\,.
\end{aligned}
\end{equation}
This expansion corresponds to the solution given in \eqref{weakKerrNNLO}.
\end{enumerate}


\section{Hartle--Thorne-type quadrupolar branch with Kerr-type spin-squared corrections} \label{Mixed}
As discussed in Section~\ref{mixedsec}, the NNLO field equations admit the nontrivial mixed solution given in \eqref{weakmixed}:
\begin{equation}
\begin{aligned}
&N=1,\quad M= -\frac{Gm}{r}+ \frac{GQP_2(\cos{\th})}{r^3}\,,\quad Z_\phi=-\frac{2GJ}{r}\sin^2{\th}\,,\\
&\g_{ij}=\text{\rm diag}\left( 1, r^2, r^2\sin^2{\th} \right)\,,\\
&\b_{rr}=\frac{2Gm}{r}-\frac{2GQ P_2(\cos{\th})}{r^3} - \frac{J^2}{m^2\,r^2} \sin^2{\th}\,,\\
&\b_{\th\th}=-\frac{2 GQP_2(\cos{\th})}{r} + \frac{J^2}{m^2} \cos^2{\th}\,,\\
&\b_{\phi\phi}=-\frac{2 GQP_2(\cos{\th})}{r}\,\sin^2{\th}  + \frac{J^2}{m^2} \sin^2{\th}\,.
\end{aligned}
\end{equation}
This solution corresponds to the weak-field branch and combines the quadrupole sector of the Hartle--Thorne geometry with the $J^2$ sector of the Kerr geometry. Therefore, one can uplift this ADM form to a relativistic metric of the form
\bea 
\frac{g_{tt}}{c^2}&=&-1 +c^2\left(\frac{2G m}{r}-\frac{2GQ P_2(\cos{\th})}{r^3} \right)-c^4\frac{2G^2mQ P_2(\cos{\th})}{r^4}\,, \nn \\
g_{t\phi}&=& -c^4\frac{2GJ\sin^2{\th}}{r}\,, \nn \\
g_{rr}&=&1+c^2\left( \frac{2G m}{r}-\frac{2GQP_2(\cos{\th})}{r^3} \right)- \frac{c^2\, J^2}{m^2\,r^2} \sin^2{\th}+c^4\frac{2G^2}{r^4}\left( 2m^2 r^2-5mQP_2(\cos{\th}) \right)  \,, \nn \\
g_{\th\th}&=& r^2-c^2\frac{2 GQP_2(\cos{\th})}{r} + \frac{c^2\, J^2}{m^2} \cos^2{\th}-c^4\frac{5 G^2 m  Q P_2(\cos{\th})}{r^2} \,, \nn \\
g_{\phi\phi}&=& r^2\sin^2{\th}-c^2\frac{2 GQP_2(\cos{\th})}{r}\sin^2{\th}+ \frac{c^2\, J^2}{m^2} \sin^2{\th}-c^4\frac{5 G^2 m  Q P_2(\cos{\th})}{r^2}\sin^2{\th} \,. 
\eea 
The weak-field NNLO branch combines an independent Hartle--Thorne-type quadrupole deformation with Kerr-like spin-squared terms. It is therefore more naturally interpreted as belonging to the broader class of Kerr-like or quasi-Kerr quadrupolar geometries, rather than as a deformation within the pure Kerr family. In particular, its weak-field structure is closer to the Kerr-like quadrupolar expansion of \cite{Frutos-Alfaro:2015gaa} than to the standard Hartle--Thorne metric, although we do not claim full equivalence with the full post-linear truncation considered in \cite{Frutos-Alfaro:2015gaa,Glampedakis:2005cf}. At the same time, the ansatz remains closely related to the Hartle--Thorne exterior metric, with the difference lying mainly in the organization of the parametrization.

\section{C-metric} \label{cmetric}
The C-metric, together with its rotating generalization, provides an exact solution of the vacuum Einstein equations \cite{Griffiths:2006tk,Hong:2004dm}. It describes one or two accelerating black holes. The metric takes the form
\begin{equation}
ds^2=\frac{1}{\Omega^2}\left[
-\frac{\mathcal Q}{\rho^2}\Big(c\,dt-a\sin^2\theta\,d\phi\Big)^2
+\frac{\rho^2}{\mathcal Q}\,dr^2
+\frac{\rho^2}{\mathcal P}\,d\theta^2
+\frac{\mathcal P\sin^2\theta}{\rho^2}
\Big((r^2+a^2)d\phi-a\,c\,dt\Big)^2
\right],\nn 
\end{equation}
with
\begin{equation}
\Omega=1+\alpha r\cos\theta,
\qquad
\rho^2=r^2+a^2\cos^2\theta, \nn 
\end{equation}
\begin{equation}
\mathcal P
=
1+\frac{2\alpha Gm}{c^2}\cos\theta
+\alpha^2 a^2\cos^2\theta, \nn 
\end{equation}
\begin{equation}
\mathcal Q
=
(1-\alpha^2 r^2)\left(r^2-\frac{2Gm}{c^2}r+a^2\right).
\end{equation}

\paragraph{Non-rotating C-metric:}
Setting $a=0$, one obtains the non-rotating C-metric background
\begin{equation}
ds^2
=
\frac{1}{\Omega^2}
\left(
- Q\,c^2\, dt^2
+\frac{dr^2}{Q}
+\frac{r^2}{P}\,d\theta^2
+P r^2 \sin^2\theta\,d\phi^2
\right),
\end{equation}
with
\begin{equation}
P =\mathcal P|_{a=0}
\qquad
Q = r^{-2}\,\mathcal Q|_{a=0}\,. 
\end{equation}
We do not present the explicit small-$c$ expansion of the C-metric in either the strong- or weak-field regime, since the resulting expressions are rather lengthy.


\bibliographystyle{JHEP}
\bibliography{bsld}
\end{document}